\documentclass[a4paper,11pt]{article}
\pdfoutput=1 

\usepackage{jheppub} 

\usepackage[T1]{fontenc} 

\usepackage{graphicx,epsfig}
\usepackage{xspace}
\usepackage{verbatim}
\usepackage{textcomp}

\def\etmiss{\ensuremath{E_{\mathrm{T}}^{\mathrm{miss}}}\xspace}

\def\ptmiss{\ensuremath{\vec p^{\mathrm{\ miss}}_\mathrm{T}\xspace}}

\newcommand{\mttwo}{\ensuremath{m_{\mathrm{T2}}}\xspace}

\def\ttbar{\ensuremath{t\bar{t}}}

\def\TeV{\ifmmode {\mathrm{\ Te\kern -0.1em V}}\else
                   \textrm{Te\kern -0.1em V}\fi}%
\def\GeV{\ifmmode {\mathrm{\ Ge\kern -0.1em V}}\else
                   \textrm{Ge\kern -0.1em V}\fi}%
                   
\def\bm#1{\mbox{\boldmath$#1$\unboldmath}}       

\def \beq{\begin{equation}}
\def \eeq{\end{equation}}
\def \bea{\begin{eqnarray}}
\def \eea{\end{eqnarray}}       

\def\Mmed{M}
\def\Msca{M_\phi}
\def\Mpse{M_a}            

\title{Determining the CP nature of spin-0 mediators in associated production of dark matter and $\bm{t\bar{t}}$ pairs}

\author[1,4]{Ulrich Haisch,}
\author[2,5]{Priscilla Pani}
\author[3,5]{and Giacomo Polesello}

\affiliation[1]{Rudolf Peierls Centre for Theoretical Physics, University of Oxford, \\ OX1 3NP Oxford, United Kingdom}
\affiliation[2]{Stockholm University - Department of Physics\\
AlbaNova University Center, 106 91 Stockholm, Sweden}
\affiliation[3]{INFN, Sezione di Pavia\\ Via Bassi 6, 27100 Pavia, Italy}
\affiliation[4]{CERN, Theoretical Physics Department, \\ CH-1211 Geneva 23, Switzerland}
\affiliation[5]{CERN, Experimental Physics Department, \\ CH-1211 Geneva 23, Switzerland}

\emailAdd{ulrich.haisch@physics.ox.ac.uk}
\emailAdd{priscilla.pani@cern.ch}
\emailAdd{giacomo.polesello@cern.ch}

\abstract{In the framework of spin-0 $s$-channel simplified models, we explore the possibility of assessing the structure of dark matter interactions through the associate production of dark matter and $t\bar{t}$ pairs. To this purpose, final states with two leptons are considered and the kinematic properties of the dilepton system is studied.  We develop a realistic analysis strategy and provide a detailed evaluation of the achievable sensitivity for the dark matter signal assuming integrated luminosities of $300 \, {\rm fb}^{-1}$  and $3 \, {\rm ab}^{-1}$ at the 14 TeV LHC. Furthermore, upper limits on the mediator masses for which the two different CP hypotheses  can be distinguished are derived. The obtained  limits on the signal strengths are finally translated into constraints on the parameter space of two spin-0 simplified models including a scenario with an extended Higgs sector.}

\preprint{CERN-TH-2016-243}

\begin{document} 
\maketitle
\flushbottom

\section{Introduction}
\label{sec:introduction}

Searches for dark matter (DM) particles constitute a key part of the physics programme at the LHC. The minimal experimental signature of DM production at a hadron collider consists of an excess of events with a final-state object $X$ recoiling against large amounts of missing transverse  energy (\etmiss). In  LHC Run I and II, the ATLAS and CMS collaborations have examined a variety of such mono-$X$ signatures involving jets of hadrons, photons, electroweak (EW) gauge bosons, top and bottom quarks as well as the Higgs boson in the final state. The cross section limits obtained from these \etmiss searches have been  interpreted in the context of  three different classes of theories: ultraviolet complete models, effective field theories~\cite{Beltran:2010ww,Goodman:2010yf,Bai:2010hh,Goodman:2010ku,Fox:2011pm,Rajaraman:2011wf} and to an increasing degree also simplified models~\cite{Bai:2010hh,Fox:2011pm,Dudas:2009uq,Goodman:2011jq,An:2012va,Frandsen:2012rk,Buchmueller:2013dya,Chang:2013oia,An:2013xka,Bai:2013iqa,DiFranzo:2013vra,Alves:2013tqa,Papucci:2014iwa,Abdallah:2014hon,Abdallah:2015ter,Abercrombie:2015wmb}. In this article, we focus on the third class of theories which has the advantage of introducing only a minimal number of additional parameters compared to the~SM, while evading  the limitations that DM effective field theories face at the LHC. Specifically we will study simplified scalar and pseudoscalar mediator models which allow for $s$-channel~DM pair production. 

New spin-0 DM mediators coupling to the SM fermions necessarily carry SM flavour quantum numbers or conversely break the SM flavour symmetry. Extra sources of flavour breaking are however severely restricted  since most  flavour measurements agree well with the corresponding SM predictions. The simplest explanation of  this empirical fact is to assume that the minimal flavour breaking consistent with the observed fermion hierarchy is also realised beyond the SM. This assumption is known as the minimal flavour violation~(MFV) hypothesis \cite{D'Ambrosio:2002ex} and implies that the couplings between any new neutral spin-0 state and SM matter are proportional to the fermion masses. One is thus naturally led to consider DM spin-0 mediators that couple most strongly to the third generation. Similar to the~SM Higgs, such resonances can be produced at the LHC  through loop-induced gluon fusion or in association with top (or bottom) quarks before they decay to the heaviest kinematically allowed SM final state or DM. 

The two main channels that have been used up to now at the LHC to search for scalars and pseudoscalars with large invisible decay widths are  $pp \to {\rm jets} + \etmiss$~\cite{Haisch:2012kf,Fox:2012ru,Haisch:2013fla,Haisch:2013ata,Buckley:2014fba,Harris:2014hga,Haisch:2015ioa,Backovic:2015soa} and  $pp \to \ttbar + \etmiss$~\cite{Buckley:2014fba,Haisch:2015ioa,Backovic:2015soa,Cheung:2010zf,Lin:2013sca,Aad:2014vea,Khachatryan:2015nua,Buckley:2015ctj}. While for realistic cuts the signal rate is  larger in the first case,  one needs to study two-jet correlations~\cite{Haisch:2013fla,Cotta:2012nj,Crivellin:2015wva}  to gain information on the structure of the portal interactions. Associated production of DM and~$\ttbar$ pairs  is also sensitive to the nature of the mediator couplings through the kinematics of the top quarks,  and through their polarisation which is accessible by studying the top-decay products.  The fully leptonic channel with both $W$ from $t \to bW$ decaying to $\ell \nu_\ell$ has compared to  $pp \to jj + \etmiss$ the advantage of a  clean final state, and the kinematic correlations between the two charged leptons can be used as a CP analyser of the underlying $pp \to \ttbar + \etmiss$ process. An interesting variable that has been proposed for separating $\ttbar$ production in the SM from new physics is the distance of the two leptons in the plane transverse to the beam direction~(azimuthal plane)~$\Delta\phi_{\ell\ell}$. This observable has been studied in some detail in the recent paper \cite{Buckley:2015ctj}  and  found to have promising discriminating power to disentangle scalar from pseudoscalar mediators.\footnote{Other observables such as $\Delta \eta_{\ell \ell}$ and $\Delta \eta_{b\bar b}$  \cite{HaischDMLHC14,Haisch:2015ioa} as well as $m_{\ttbar}$ \cite{Backovic:2015soa} have been proposed as probes of the CP nature of the DM $\ttbar$  interactions, but a detailed evaluation of the achievable sensitivity for the DM signal has not been performed in these cases.}

In  this work an  alternative search strategy is devised which exploits the properties of $\cos\theta_{\ell\ell}\equiv\tanh \left (\Delta\eta_{\ell\ell}/2 \right )$ where  $\Delta \eta_{\ell\ell}$  is the  difference in pseudorapidity of the two charged leptons. This variable has been introduced in \cite{Barr:2005dz} as a proxy of the polar angle for the production of a pair of hypothetical particles both of which decay into a visible and an invisible particle.  In the case of the $pp \to \ttbar + \etmiss$ signal, we find that  in contrast to the~$\Delta\phi_{\ell\ell}$ distribution used  in \cite{Buckley:2015ctj}, the~$\cos\theta_{\ell\ell}$ distribution is affected only in a minor way by the selections in the transverse plane that are needed to suppress the dominant SM background from top-quark final states to manageable levels. The variable $\cos\theta_{\ell\ell}$ is thus a promising observable when analysing~$\ttbar + \etmiss$ events since it provides good CP-discriminating power also when experimental selection cuts are employed. Based on it we develop a realistic strategy for the detection of the DM signal at future LHC runs and determine the exclusion limits on the signal models using both a simple cut-and-count experiment as well as a shape-fit analysis. For the selection criteria adopted in this study the discovery reach in the $\ttbar + \etmiss$ channel depends sensitively on the systematic uncertainty assigned to the estimate of SM backgrounds, which are dominated by $\ttbar Z$.  A~good experimental understanding of this process is hence a prerequisite to exploit the large data sets that the LHC is expected to provide at  high luminosity  (HL-LHC). We~also derive upper limits on the mediator masses for which the two different CP hypotheses  can be distinguished and translate the obtained model-independent  limits on the signal strengths into constraints on the parameter space of two specific simplified DM models. 

This article is structured as follows. In Section~\ref{sec:simplified} we discuss the relevant spin-0 interactions, while  the basic properties  of the $\ttbar + \etmiss$ signal are studied in Section~\ref{sec:anatomy}. A~description of  our Monte Carlo (MC) simulations is presented in Section~\ref{sec:montecarlo}. This part contains a discussion of the signal and background generation and explains how  physics objects are built in our analysis. The actual analysis strategy is presented in Section~\ref{sec:analysis} spelling out all selection criteria and illustrating their impact on the DM signal and the SM backgrounds. In Section~\ref{sec:results} we present the numerical results of our analysis providing a detailed evaluation of the achievable sensitivity for the $\ttbar + \etmiss$ signature at future LHC runs. The constraints on the parameter space of two different spin-0 simplified~DM models are also derived in this section.  Our conclusions and a brief outlook are given in Section~\ref{sec:conclusions}. 

\section{Simplified models}
\label{sec:simplified}

The simplified DM models we are considering in our article can be described by the following interactions
\beq \label{eq:lagrangians}
\begin{split}
{\cal L}_\phi & \supset -g_\chi \phi \bar \chi \chi - \frac{\phi}{\sqrt{2}} \sum_{q=u,d,s,c,b,t}  g_q y_q \bar q q \,, \\[2mm]
{\cal L}_a & \supset -i g_\chi a \bar \chi \gamma_5 \chi - i  \hspace{0.25mm} \frac{a}{\sqrt{2}} \sum_{q=u,d,s,c,b,t}  g_q y_q \bar q \gamma_5 q \,. 
\end{split}
\eeq
Here $\phi$ is a scalar while $a$ is a pseudoscalar, $\chi$ represents the DM particle  which we assume to be a Dirac fermion, $g_\chi$ is a dark sector Yukawa coupling and $y_q = \sqrt{2} m_q/v$ are the SM quark Yukawa couplings with $v \simeq 246 \, {\rm GeV}$ the Higgs vacuum expectation value (VEV). 

Notice that the simplified models (\ref{eq:lagrangians}) are valid descriptions of the physics below the EW scale as long as the new scalar $\phi$ does not mix strongly with the SM Higgs boson. In such a case the model dependence associated to the full scalar sector is captured by  the portal couplings $g_\chi$ and $g_q$. The simplest choice of quark couplings compatible with the MFV hypothesis is universal $g_q = g_v$ and realised for instance in the spin-0 simplified models recommended by the ATLAS/CMS DM Forum (DMF)~\cite{Abercrombie:2015wmb}. If the SM Higgs sector is extended to a two Higgs doublet model (2HDM) other MFV coupling patterns are however possible. For instance within the alignment/decoupling limit of 2HDM of~type~II~(2HDMII) plus singlet extensions the relevant couplings are given by (see for example~\cite{Ipek:2014gua})
\beq \label{eq:2HDM}
g_\chi = y_\chi \cos \theta \,, \qquad 
g_{u,c,t} = \pm \sin \theta \cot \beta \,, \qquad  
g_{d,s,b} =  -\sin \theta \tan \beta \,.
\eeq 
Here $y_\chi$ denotes the coupling of the singlet to DM and $\theta$ is the mixing angle between the singlet and the heavy CP-even or CP-odd Higgs state and the $+$ ($-$) sign in $g_{u,c,t}$ holds for a scalar (pseudoscalar). Furthermore $\tan \beta$ represents the ratio of VEVs of the two Higgs doublets and for small $\tan \beta$ one obtains the hierarchy $g_{t} y_t \gg g_{q} y_q$ with $q = u,d,s,c,b$.
 
\section{Anatomy of the signal}
\label{sec:anatomy}

\begin{figure}[!t]
\begin{center}
\includegraphics[width=0.49\textwidth]{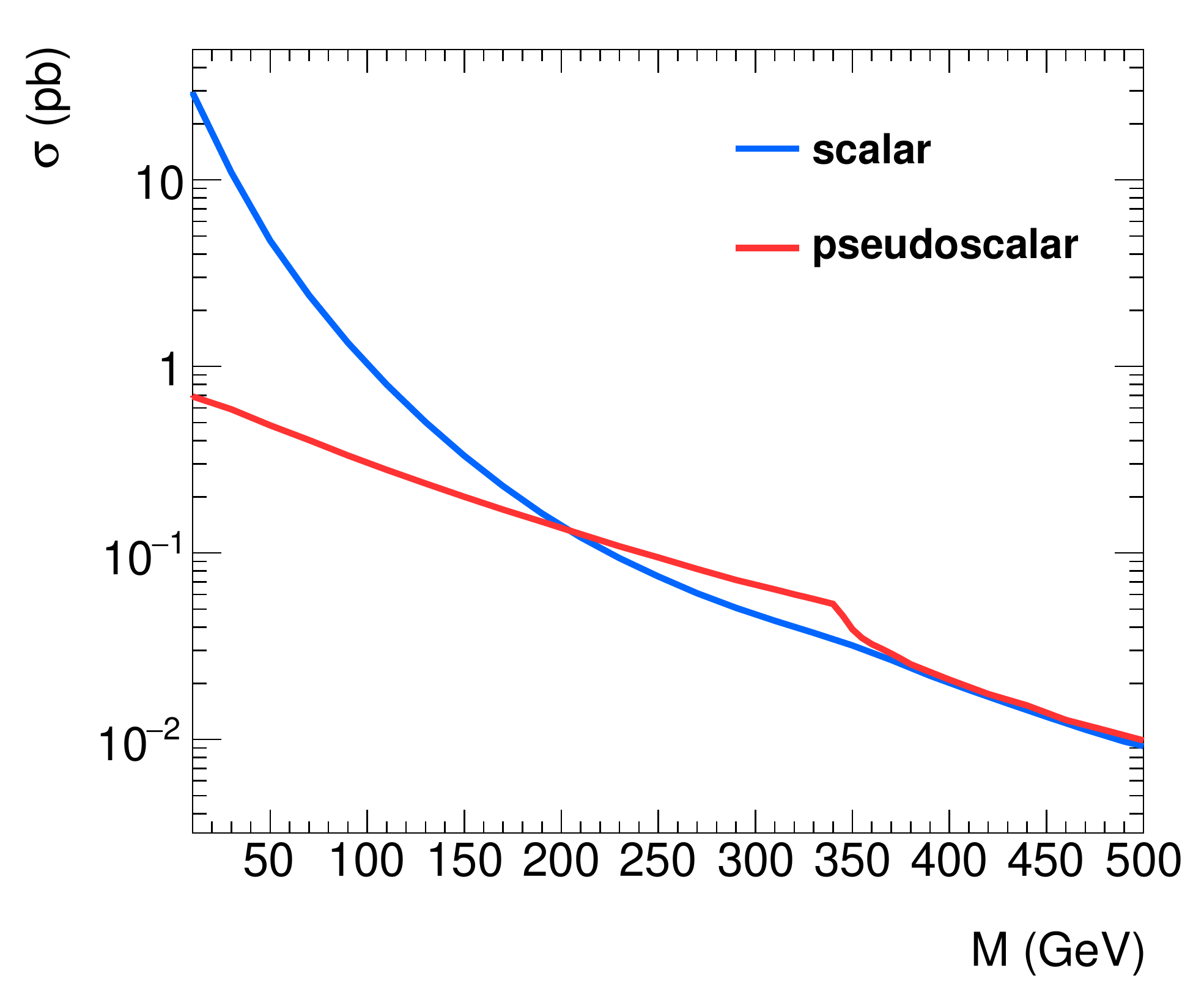}  
\includegraphics[width=0.49\textwidth]{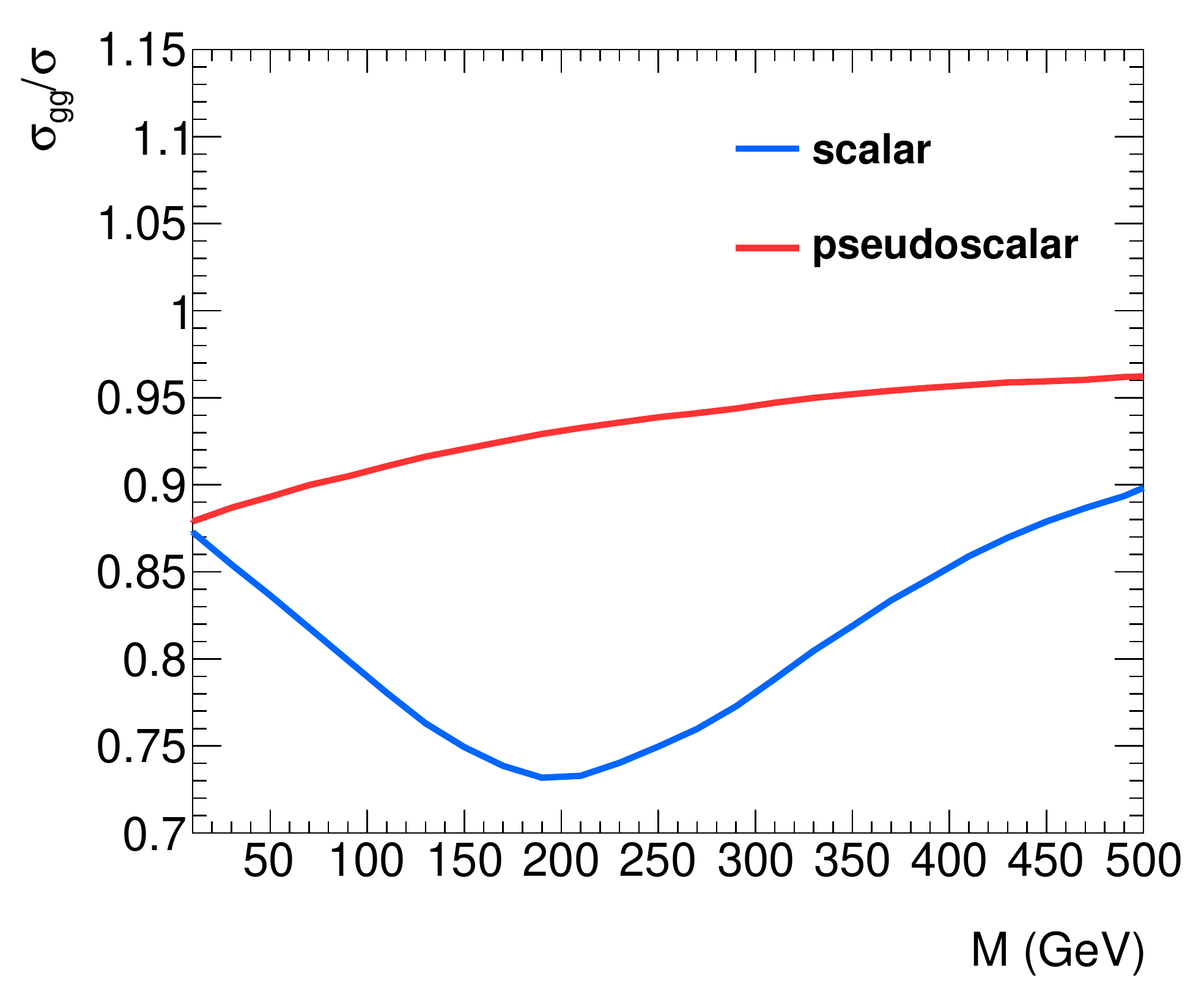}  
\vspace{2mm}
\caption{\label{fig:xsec}  Left: Total production cross section for $pp \to \ttbar  + \etmiss$ as a function of the mediator mass. Right: Mediator mass dependence of the ratio of gluon-fusion production rate to the total production cross section. Both panels correspond to $\sqrt{s}=14 \, {\rm TeV}$, employ $m_\chi = 1 \, {\rm GeV}$ and $g_\chi = g_t =1$~and assume a minimal decay width for the mediator. The predictions for a scalar~(pseudoscalar) mediator are shown in blue (red).}
\end{center}
\end{figure}

In Figure~\ref{fig:xsec} we show the production cross section for $pp \to \ttbar  + \etmiss$ (left panel) and the corresponding fraction of events arising from gluon fusion (right panel) for a centre-of-mass energy ($\sqrt{s}$) of $14 \, {\rm TeV}$. The displayed results have been obtained at next-to-leading order~(NLO) with the help of {\tt  MadGraph5\_aMC@NLO}~\cite{Alwall:2014hca} employing the {\tt DMsimp} implementation~\cite{Backovic:2015soa} of the simplified models (\ref{eq:lagrangians}) and {\tt NNPDF3.0}~parton distribution functions~(PDFs)~\cite{Ball:2014uwa}. From the left panel one observes that for very low mediator masses $\Mmed = \Msca$ or $\Mpse$ the cross section associated to scalar exchange (blue curve) is larger than that for a pseudoscalar (red curve)  by more than an order of magnitude. At around $M \simeq 200 \, {\rm GeV}$ the two  predictions then become alike, while at higher masses the rate for pseudoscalar production is always slightly larger than that for a scalar. In the right plot, one sees  that at the LHC  the gluon-fusion channel is the dominant production mode independently of the~CP nature of the mediator and amounts to   roughly~$85\%$ of the total cross section  for $\Mmed \simeq 10 \, {\rm GeV}$.  The functional dependence of $\sigma_{gg}/\sigma$ is however  different in the two cases. While in the CP-even case the fraction of gluon-fusion initiated events first decreases until about $\Mmed \simeq 200 \, {\rm GeV}$ and then starts rising, in the case of the CP-odd mediator the ratio~$\sigma_{gg}/\sigma$ is a steadily increasing function of $M$. 

The features observed in Figure~\ref{fig:xsec} can be  understood qualitatively in terms of two physical effects~\cite{Backovic:2015soa}. The first effect is related to the fact that a spin-0 state which has a mass much lighter than all of the relevant energy scales in a process $pp \to X$ can be treated as a parton which is radiated off the individual particles in the final state $X$. The process $pp \to \ttbar + \phi/a \; (\phi/a \to \chi \bar \chi)$ can thus be thought as $pp \to \ttbar$ followed by the radiation of $\phi/a$ from the final-state heavy quark lines with a subsequent decay of the spin-0 mediator to DM. This procedure is guaranteed to correctly reproduce the collinear divergencies associated with the emission of a massless $\phi/a$ state. The observed radiation pattern is determined by the leading (universal) fragmentation function $f_{t \to \phi/a} (x)$ which take the form  \cite{Dawson:1997im, Dittmaier:2000tc}
\beq \label{eq:ffs}
\begin{split}
f_{t \to \phi} (x) & = \frac{g_t^2}{(4 \pi)^2} \left [ \frac{4 \left (1 -x \right )}{x} + x \ln \left ( \frac{s}{m_t^2} \right ) \right ] \,, \\[2mm]
f_{t \to a} (x) & = \frac{g_t^2}{(4 \pi)^2} \left [  x \ln \left ( \frac{s}{m_t^2} \right ) \right ] \,,
\end{split}
\eeq
in the simplified models described by (\ref{eq:lagrangians}). These results are valid for $s \gg 4 m_t^2 \gg M^2$ and $\ln \left (s/m_t^2 \right ) \ll 1$ where $\sqrt{s} = 2 E/x$ with $E$ the energy of the emitted spin-0 particle. From~(\ref{eq:ffs}) one sees that due to the soft singularity proportional to~$1/x$ a light scalar is radiated off top quarks preferentially with small energy (or equivalent small momentum fraction $x$). The soft-enhanced term is instead absent in the case of the pseudoscalar mediator. These features explain the order of magnitude difference between the total rates of the scalar and pseudoscalar mediators for masses $\Mmed \ll 2 m_t$.

\begin{figure}[!t]
\begin{center}
\includegraphics[width=0.9\textwidth]{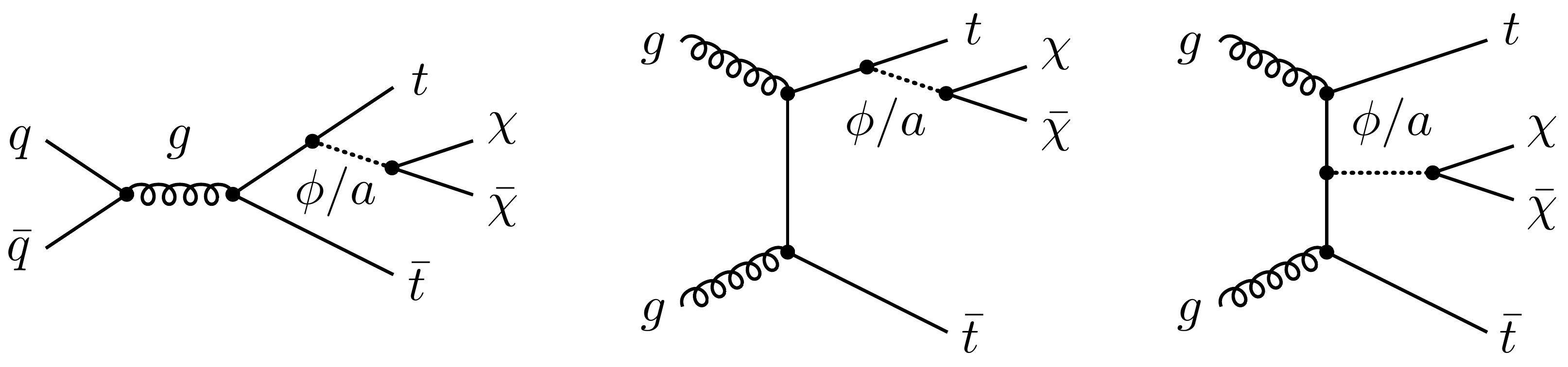}  
\vspace{2mm}
\caption{\label{fig:diagrams}  Examples of  LO diagrams that give rise to  a $\ttbar + \etmiss$ signature through the exchange of a colourless spin-0 mediator. In the quark-fusion channel (left) only  contributions from mediator fragmentation appear, while in the case of the gluon-fusion channel both mediator-fragmentation~(center) and top-fusion (right) diagrams are present.}
\end{center}
\end{figure}

The second important difference between $\sigma(pp \to \ttbar + \phi)$ and  $\sigma(pp \to \ttbar + a)$ with $\phi$ and $a$ subsequently decaying to DM  can be understood by considering the spin-averaged and colour-averaged squared matrix elements for the production of an on-shell  spin-0 state with mass $\Mmed = \sqrt{s}$ from a top-quark pair. The corresponding squared matrix elements are given by 
\beq \label{eq:squaredMEs}
\overline{\sum} \, \big | {\cal M} (\ttbar \to \phi) \big |^2 = \frac{g_t^2 s}{12}  \hspace{0.25mm} \beta^2 \,, \qquad  
\overline{\sum} \, \big |  {\cal M} (\ttbar \to a) \big |^2 = \frac{g_t^2 s}{12} \,,
\eeq
with $\beta = \sqrt{1 - 4 m_t^2/s}$ the velocity of the top quarks in the top-pair rest frame. From the above formulas one observes that close to the $\ttbar$ threshold located at $4 m_t^2$ the production of a scalar in top-fusion is compared to that of a pseudoscalar suppressed by two powers of $\beta$. It follows that in cases where either the DM pair or the mediator is produced close to threshold,  the production cross section of the pseudoscalar mediator is expected to be larger than that of a scalar. This is precisely what one observes in the left panel of Figure~\ref{fig:xsec}. As it leads to a pronounced kink in the pseudoscalar case,  the opening of the $\ttbar$ threshold is also clearly visible in this plot. The threshold suppression of $\ttbar \to \phi$ production finally explains the $M$ dependence of the ratio $\sigma_{gg}/\sigma$ with a dip at $\Mmed \simeq 200 \, {\rm GeV}$ as shown on the right in the latter figure. 

The above observations can also be used to identify which leading order (LO) diagrams give the dominant contribution to the $\ttbar + \etmiss$ signature in the case of a scalar or pseudoscalar. Representative examples of the three possible tree-level topologies are shown in Figure~\ref{fig:diagrams}. From the previous  discussion is should be clear that at the LHC the $\sigma (pp \to \ttbar + \phi \; (\phi \to \chi \bar \chi))$ cross section is dominated by the gluon-fusion graph with a mediator fragmentation shown in the centre of the latter figure. In the case of the pseudoscalar cross section $\sigma (pp \to \ttbar + a \; (a \to \chi \bar \chi))$, on the other hand,  both mediator-fragmentation and top-fusion diagrams in gluon-fusion are  relevant. The latter contribution is displayed on the right in~Figure~\ref{fig:diagrams}.

\begin{figure}[t!]
\begin{center}
\includegraphics[width=0.49\textwidth]{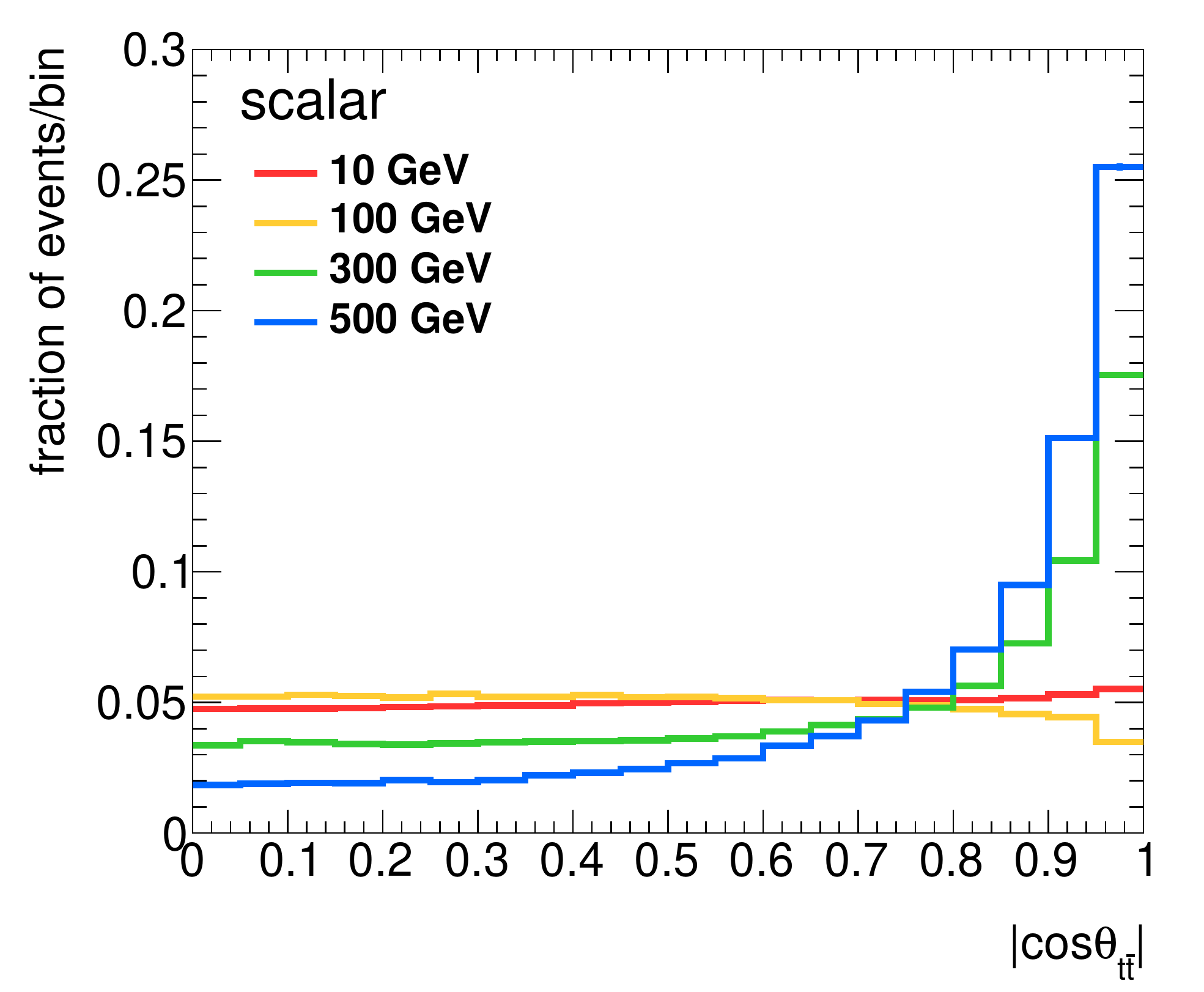}
\includegraphics[width=0.49\textwidth]{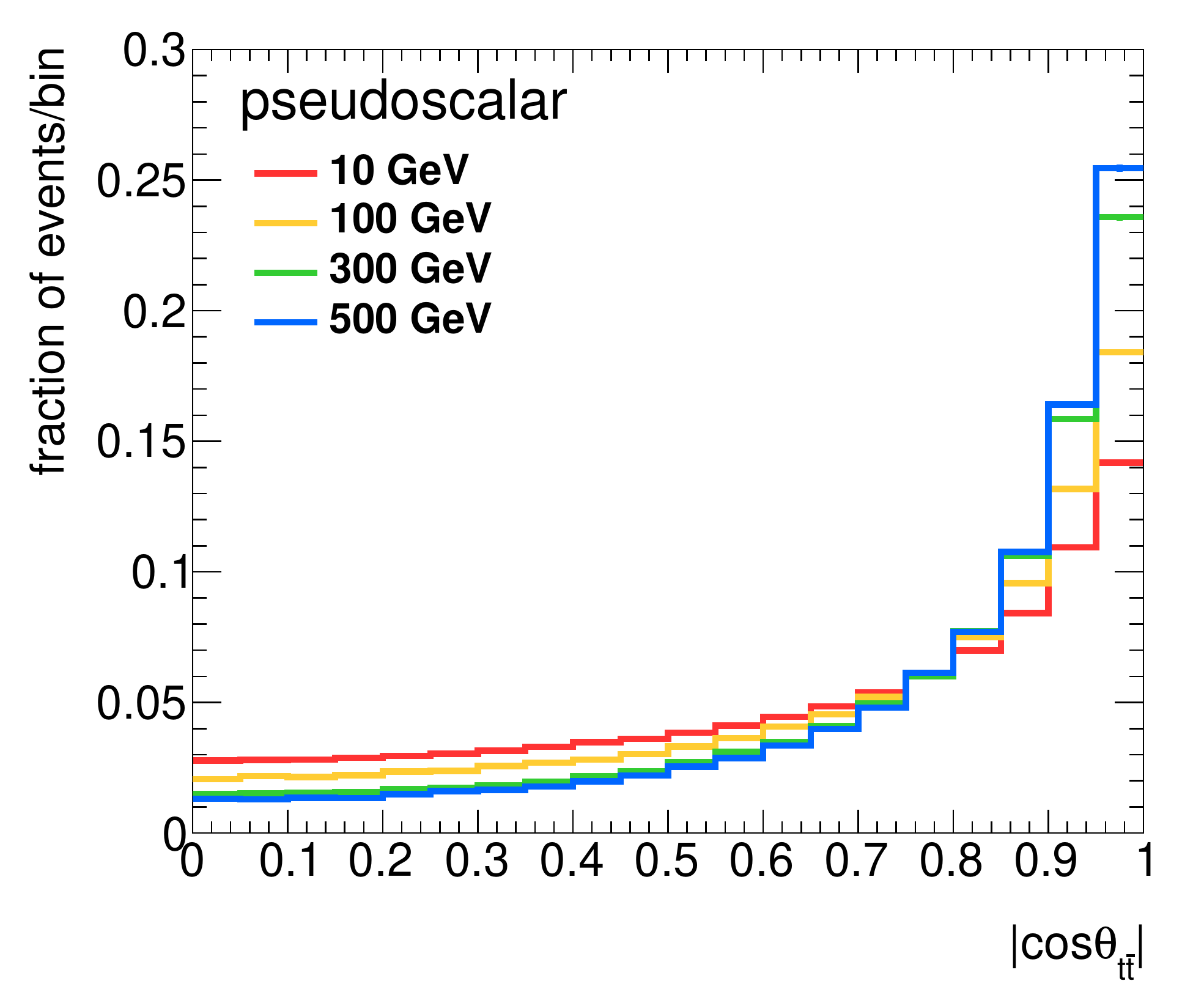}
\includegraphics[width=0.49\textwidth]{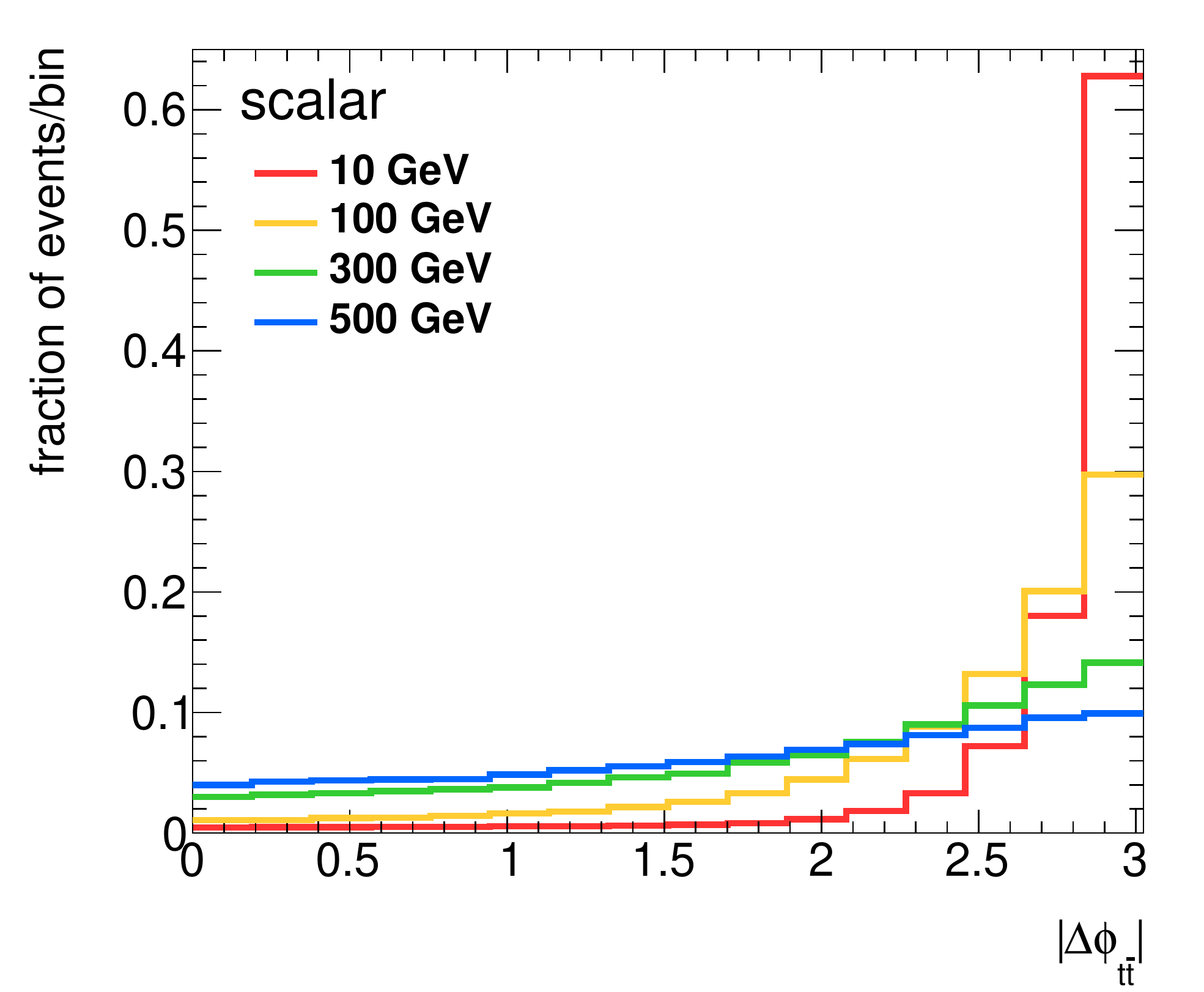}
\includegraphics[width=0.49\textwidth]{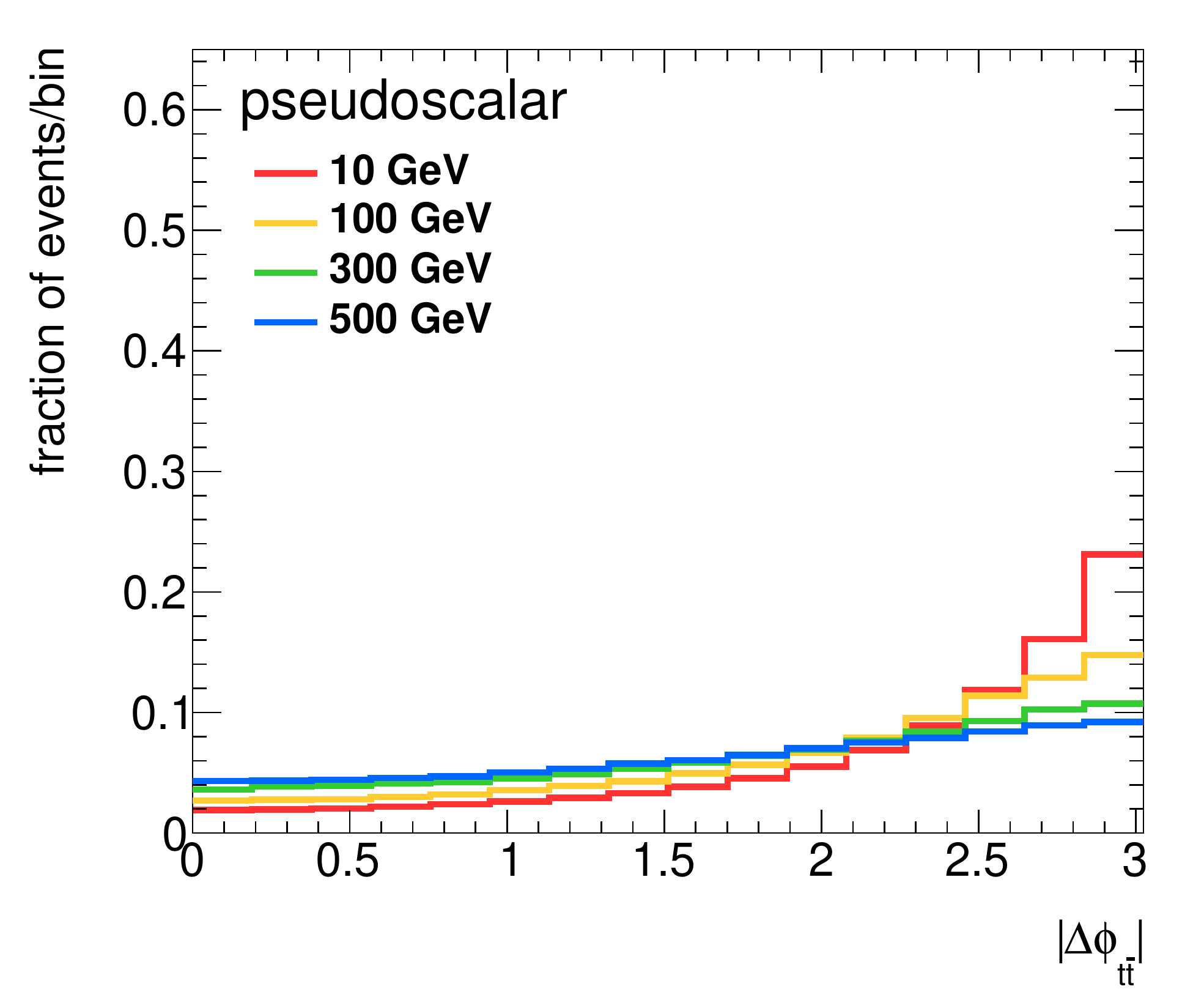}
\caption{Normalised distributions of the $|\hspace{-0.4mm} \cos\theta_{\ttbar}|$  (upper row) and $|\Delta\phi_{\ttbar}|$ (lower row)  variables for four different~scalar (left) and  pseudoscalar (right) benchmark models. The red, yellow, green and blue curves correspond to mediator masses of $10 \, {\rm GeV}$, $100\, {\rm GeV}$, $300\, {\rm GeV}$ and $500 \, {\rm GeV}$, respectively.}
\label{fig:ttangular}
\end{center}
\end{figure}

In the case of $\ttbar + \etmiss$  production information on the CP nature of the coupling between the mediator and  top quarks is encoded in the correlations between the final-state top quarks and their decay products. The two variables that we will study in this section are the  $\cos \theta_{\ttbar} \equiv \tanh \left ( \Delta \eta_{\ttbar}/2 \right )$  variable and the azimuthal angle difference~$\Delta \phi_{\ttbar}$ of the $\ttbar$ system. In Figure~\ref{fig:ttangular} we  present NLO predictions for the $| \hspace{-0.4mm} \cos \theta_{\ttbar}|$ and $|\Delta \phi_{\ttbar}|$ distributions for four different realisations of the scalar and pseudoscalar  models introduced in (\ref{eq:lagrangians}). The shown spectra are all normalised to unity. Comparing the two panels in the upper row of the figure, one observes that for a light mediator of $10 \, {\rm GeV}$~(red curves) and~$100 \, {\rm GeV}$~(yellow curves) the  normalised~$| \hspace{-0.4mm} \cos \theta_{\ttbar}|$ distributions are almost flat in the scalar case, while they are enhanced toward larger values of $| \hspace{-0.4mm} \cos \theta_{\ttbar}|$ if the mediator is a pseudoscalar. This feature can be understood by recalling that as a result of the soft singularity of the fragmentation function $f_{t \to \phi} (x)$ in (\ref{eq:ffs}) emissions of soft~$\phi$ fields are favoured. Compared to a pseudoscalar, a scalar  of the same mass hence tends to be produced more forward, which in turn means that the accompanying top quarks are produced more central. In consequence the difference in pseudorapidity of the top-antitop pair  $\Delta \eta_{\ttbar}$ and likewise~$\cos \theta_{\ttbar}$ are expected to be smaller on average for a light scalar than a pseudoscalar. For a heavier mediator of mass~$300 \, {\rm GeV}$~(green curves) and $500 \, {\rm GeV}$~(blue curves) the soft enhancement of scalar fragmentation is no longer active so that the $| \hspace{-0.4mm} \cos \theta_{\ttbar}|$ spectra are  peaked at 1 irrespectively of the CP nature of the mediator. 

Similar arguments can be used to qualitatively explain the results for the $|\Delta \phi_{\ttbar}|$ distributions as shown  in the lower panels of Figure~\ref{fig:ttangular}. One first notices that all spectra peak at the maximum allowed angle of $\pi$. However, for pseudoscalar mediators the distributions at large $|\Delta \phi_{\ttbar}|$ values are flatter in comparison to the scalar cases, in particular for the low mass benchmarks. This can be understood as follows.  In the CP-even case  soft emissions are favoured, and consequentially the top and antitop quarks prefer to fly in opposite directions in the transverse plane, which corresponds to large $|\Delta \phi_{\ttbar}|$ values. Because relative to the scalar case the emission of a pseudoscalar tends to be harder, for a CP-odd state  the top quark and the mediator instead  end up in  a back-to-back position in the transverse plane and as a result the $|\Delta \phi_{\ttbar}|$ spectrum flattens. The impact of soft effects is  less pronounced for heavier mediators for which the $|\Delta \phi_{\ttbar}|$ distributions become all very similar. 

The above discussion should have made clear that the $\ttbar$ system can be used as an analyser of the CP properties of the mediation mechanism in $\ttbar + \etmiss$ production. Given the presence of four invisible particles in the dilepton final state, the directions of the two top quarks are however experimentally not directly accessible. Indirection information on the relative orientation of the top quarks has hence to be obtained from measurements of the angular distributions of the charged lepton pairs resulting from top decays. The question is whether the discriminating power of angular variables such as $\cos\theta_{\ell\ell}$  and $\Delta\phi_{\ell\ell}$ survive the experimental cuts necessary to extract the DM signal from the SM backgrounds, and for which range of mediator masses and couplings it will be possible to perform the needed measurements with the foreseen integrated luminosity of the LHC. These issues will be addressed in the following two sections through detailed MC analyses.

\section{MC simulations}
\label{sec:montecarlo}

In this section we provide a brief description of the MC simulations used to generate both the DM signal and the SM backgrounds and explain how electrons,  muons, photons, jets and \etmiss are built  in our detector simulation. Throughout our analysis we will consider~$pp$ collisions at $\sqrt{s} = 14 \, {\rm TeV}$. 

\subsection{Signal generation}
\label{sec:signalgeneration}

As in Section~\ref{sec:anatomy} our signal samples are generated at NLO using  the {\tt DMsimp} simplified model implementation together with {\tt  MadGraph5\_aMC@NLO} and {\tt NNPDF3.0} PDFs. The final-state top quarks and $W$ bosons are decayed  with {\tt MadSpin}~\cite{Artoisenet:2012st} and the events are showered with {\tt PYTHIA~8.2} \cite{Sjostrand:2014zea} using the {\tt FxFx} NLO jet matching prescription~\cite{Frederix:2012ps}. We consider five different values of the mediator mass $\Mmed$, varying from $10 \, {\rm GeV}$ to $500\, {\rm GeV}$ for both the case of a scalar and a pseudoscalar mediator.  The mass of the DM particles is set to $m_\chi = 1 \, {\rm GeV}$ and we employ $g_\chi = g_t =1$ for the mediator couplings  to tops and to DM. The width of the mediator is assumed to be minimal and calculated at tree level using {\tt  MadGraph5\_aMC@NLO}. Since in the narrow width approximation the signal prediction factorises into the cross section for $pp \to \ttbar + \phi/a$ production times the $\phi/a \to \chi \bar \chi$ branching ratio, changing the mediator width only leads to a rescaling of the signal strength. The experimental acceptance is instead insensitive to the mediator width, and therefore it is sufficient to generate samples for a single  coupling choice.  The predictions for other couplings values can then be simply obtained by scaling with the corresponding invisible branching ratio of the mediator.

\subsection{Background generation}
\label{sec:backgroundgeneration}

In order to describe the $\ttbar + \etmiss$ backgrounds accurately,  SM processes involving at least two leptons coming from the decay of  vector bosons are generated. Backgrounds either with fake electrons from jet misidentification or with real non-isolated leptons  from the decay of heavy flavours are not considered in our analysis, as a reliable estimate of these backgrounds would require a simulation of detector effects beyond the scope of this work.\footnote{The ATLAS  analysis \cite{Aad:2014qaa} employs kinematic variables similar to the ones used in our study and finds that in the relevant signal region the background from non-prompt leptons is negligible. We take this as a strong indication that the background from non-prompt leptons is also not an issue in our case.} The backgrounds from $\ttbar$~\cite{Campbell:2014kua}, $tW$~\cite{Re:2010bp}, $WW$, $WZ$ and $ZZ$ production~\cite{Melia:2011tj,Nason:2013ydw} were all generate at NLO with {\tt POWHEG~BOX} \cite{Alioli:2010xd}. The ${\rm jets}+Z$ samples are generated at~LO with {\tt  MadGraph5\_aMC@NLO} and contain up to four jets. {\tt  MadGraph5\_aMC@NLO} is also used to simulate the $\ttbar V$ backgrounds with $V = W,Z$ at LO with a multiplicity of up to two jets. All partonic events are showered with {\tt PYTHIA~8.2}. The samples produced with {\tt POWHEG~BOX} are normalised to the NLO cross section given by the generator, except $t\bar{t}$ which is normalised to the  cross section obtained at NNLO plus next-to-next-to-leading logarithmic accuracy~\cite{Czakon:2011xx,Czakon:2013goa}. The ${\rm jets} + W/Z$ samples are normalised to the known NNLO cross sections~\cite{Anastasiou:2003ds,Gavin:2012sy}, while in the case of the~$\ttbar V$~samples  the NLO cross sections calculated with {\tt  MadGraph5\_aMC@NLO} are used as normalisations.  

\subsection{Detector smearing}
\label{sec:detectorsmearing}

Starting from the stable particles in the generator output, the following physics objects are built: electrons,  muons, photons, jets and \etmiss. Jets are constructed from the true momenta of particles interacting in the calorimeters except muons using an anti-$k_t$ algorithm~\cite{Cacciari:2008gp} with a parameter $R=0.4$, as implemented in  {\tt FastJet}~\cite{Cacciari:2011ma}. The variable \ptmiss \ with magnitude~\etmiss is defined at truth level,~i.e.~before any detector effects are applied, as  the negative of the vector sum of the transverse momenta of all the invisible particles (neutrinos and DM particles in our case). The effect of the detector on the kinematic quantities of interest is simulated by applying a Gaussian smearing to the momenta of  the different reconstructed objects and reconstruction efficiency factors. The parametrisation of the smearing  and reconstruction efficiency  as a function of momentum and pseudorapidity of the objects is tuned to mimic the performance reported by ATLAS for Run I at the~LHC~\cite{Aad:2009wy}. The discrimination of signal from background is strongly affected by the assumed experimental smearing of \etmiss, which is the main handle to tame  the enormous~$\ttbar$~background. To this purpose, the transverse momenta  of unsmeared jets, electrons and muons are subtracted from the truth \etmiss and replaced by the corresponding smeared quantities.  The residual truth imbalance  is then smeared as a function of the scalar sum of the transverse momenta of the particles not assigned to jets or electrons. 

\section{Analysis strategy}
\label{sec:analysis}

The final state targeted in our study contains two leptons, two jets with $b$-hadrons ($b$-jets) from the decay of the top quarks as well as a significant amount of  \etmiss  associated to both the DM particles and the neutrinos from the $W \to \ell \nu_\ell$ decays. 

In order to understand which  discriminators and kinematic variables are useful to separate signal and background, one first has to recall the different types of SM backgrounds that can resemble the feature of a $\ttbar +\etmiss$ signal.  In fact, the SM backgrounds can be divided into three distinct classes:
\begin{itemize}
\item[$1)$] Top backgrounds:  The final states $\ttbar$ and $tW$ with one or two top quarks fall into this class. These backgrounds are characterised by  significant jet  activity from the top decays and the presence of $b$-jets. The two neutrinos from  $W \to \ell \nu_\ell$  represent the dominant $\etmiss$ contribution.  The~$\ttbar$ and $tW$ backgrounds can mimic the signal if extra $\etmiss$ arises either from neutrinos produced in  heavy-quark decays or from   jet-momenta  mismeasurement in the detector. 

\item[$2)$] Reducible backgrounds: The second class of backgrounds  comprises events from $WW$, $WZ$, $ZZ$ and ${\rm jets} + Z$ production where jets arise from  QCD radiation. Such configurations have compared to the $\ttbar + \etmiss$ signal on average less jet activity and often no $b$-tagged jets. In addition the amount of $\etmiss$ for these backgrounds is typically small compared to that of the DM signal.

\item[$3)$]  Irreducible backgrounds: The final states $\ttbar V$  form the third class of backgrounds. Here the $\etmiss$ signature arises from the decays $W\to \ell\nu_\ell$ or  $Z\to \nu_\ell \bar \nu_\ell$. These backgrounds are irreducible in the sense that the  resulting final-state configurations of $\ttbar V$  and the $\ttbar + \etmiss$ signal are very similar. 

\end{itemize}

\begin{figure}[t!]
\begin{center}
\includegraphics[width=0.49\textwidth]{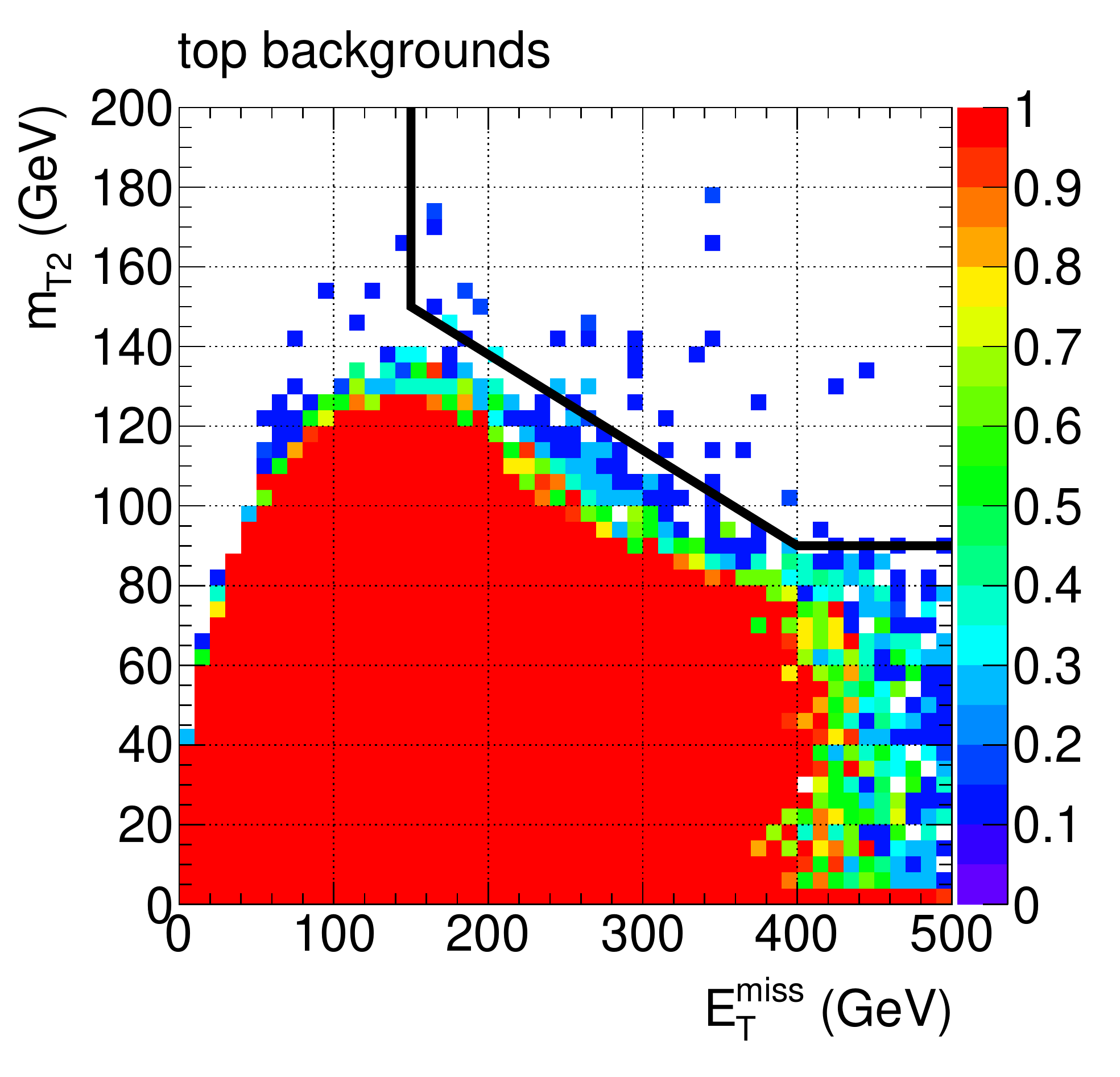}
\includegraphics[width=0.49\textwidth]{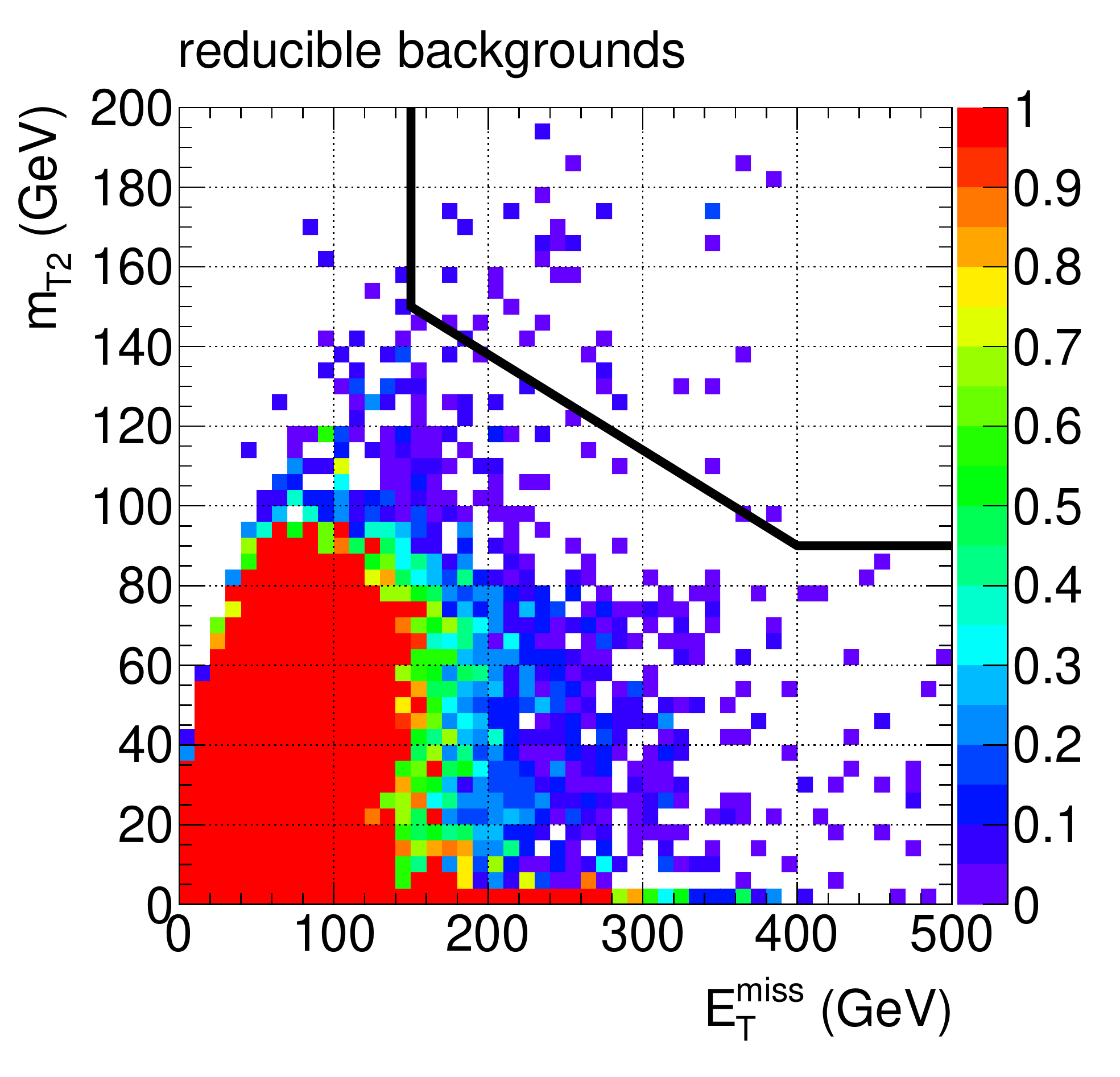}
\includegraphics[width=0.49\textwidth]{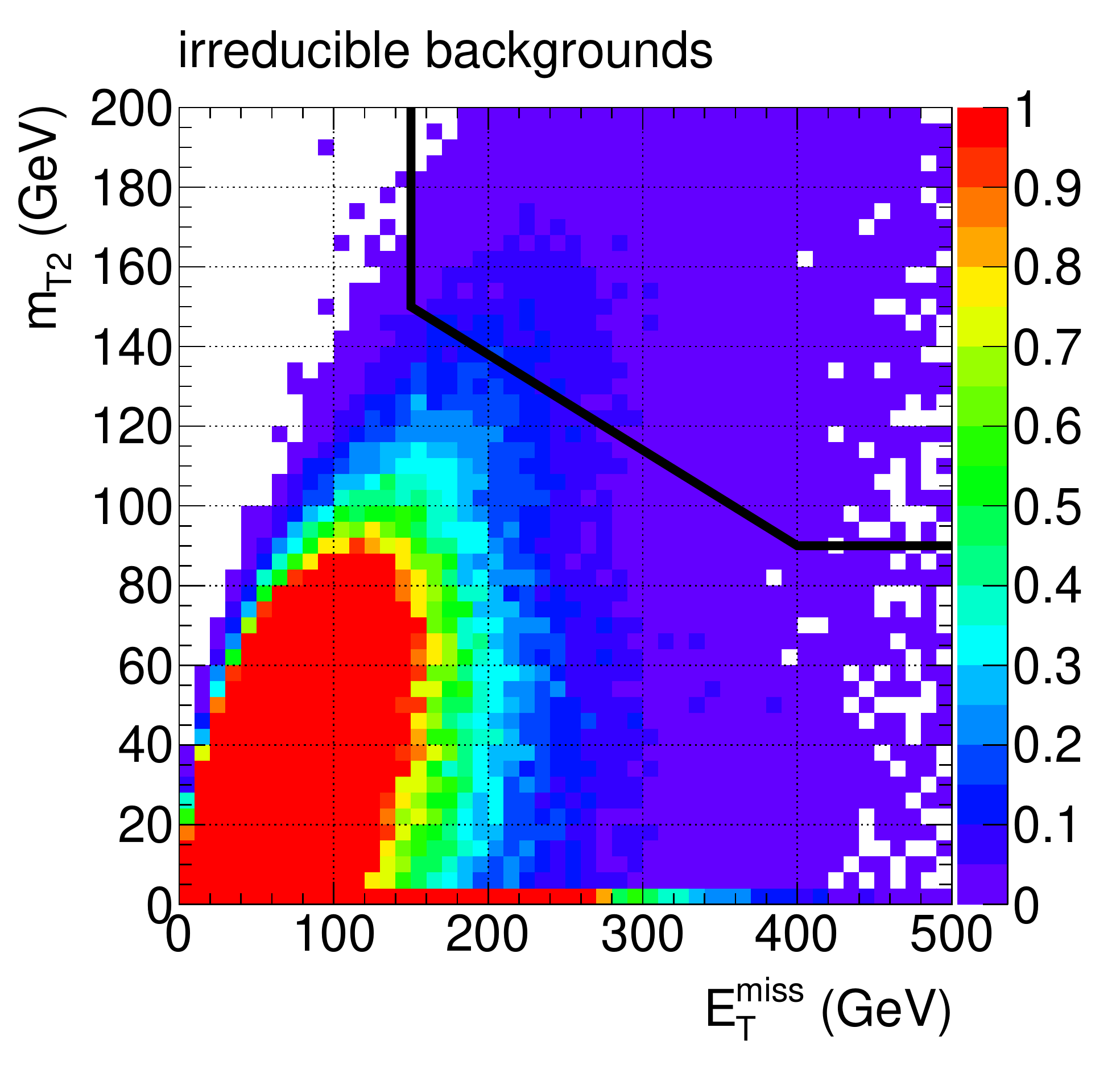}
\includegraphics[width=0.49\textwidth]{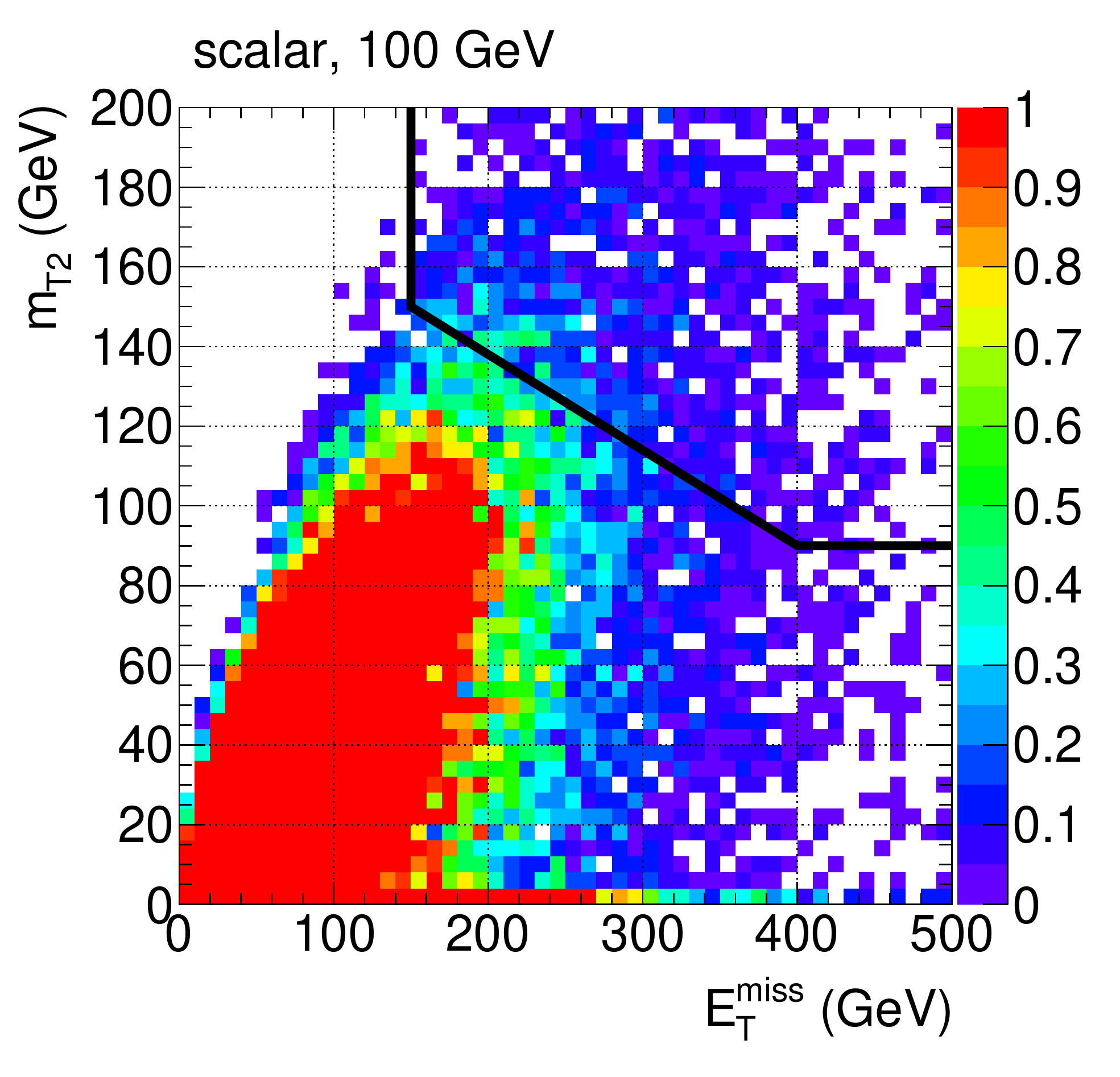}
\caption{Distributions of events in the  $\etmiss \hspace{0.1mm}$--$\,\mttwo$  plane  for the three different types of backgrounds and a signal point. The signal prediction corresponds to a scalar mediator with mass~$\Msca = 100 \, {\rm GeV}$ and assumes $m_\chi = 1 \, {\rm GeV}$ and $g_\chi = g_t = 1$. The selection cuts as defined in the text  except $\etmiss$, $\mttwo$  and $C_{\rm em}$ are imposed. The scale on the $z$ axis is saturated at $1 \, {\rm event}/{\rm bin}/(100 \, {\rm fb}^{-1})$. The area in the upper right corner above the black line represents the signal region used in our analysis.}
\label{fig:etm_mttwo}
\end{center}
\end{figure}

The natural variable for separating DM signals from SM backgrounds is $\etmiss$, which in our case turns out to be very powerful in reducing both  the top and the reducible backgrounds. Unfortunately, a selection in $\etmiss$ alone is not enough to fully tame the overwhelming top backgrounds, which exhibit long $\etmiss$   tails. A variable that is effective in suppressing all backgrounds where the two leptons are produced in the decay of two $W$ bosons is the \mttwo variable \cite{Lester:1999tx, Barr:2003rg} 
\begin{equation} \label{eq:mT2}
m^2_\mathrm{T2} (\vec p^{\ \ell_i}_\mathrm{T}, \vec p^{\ \ell_j}_\mathrm{T}, \vec p^{\mathrm{\ miss}}_\mathrm{T}) \equiv \min_{\vec q^{\ 1}_\mathrm{T} + \vec q^{\ 2}_\mathrm{T} = \vec p^{\mathrm{\ miss}}_\mathrm{T}}  \left \{ \max \left [ m^2_\mathrm{T}( \vec p^{\ \ell_i}_\mathrm{T}, \vec q^{\ 1}_\mathrm{T}),  m^2_\mathrm{T}( \vec p^{\ \ell_j}_\mathrm{T}, \vec q^{\ 2}_\mathrm{T} ) \right ] \right \} \,,
\end{equation}
which can be calculated using the momenta $\vec p^{\ \ell_i}_\mathrm{T}$ and $\vec p^{\ \ell_j}_\mathrm{T}$  of the two leptons and \ptmiss. In~(\ref{eq:mT2}) the parameter $m_\mathrm{T}$ denotes  the transverse mass and $\vec q^{\ 1}_\mathrm{T}$ and $\vec q^{\ 2}_\mathrm{T}$ are auxiliary vectors. The minimum is taken over all the possible choices of  $\vec q^{\ 1}_\mathrm{T}$ and $\vec q^{\ 2}_\mathrm{T}$ which satisfy  the equality $\vec q^{\ 1}_\mathrm{T} + \vec q^{\ 2}_\mathrm{T} = \vec p^{\mathrm{\ miss}}_\mathrm{T}$.  The \mttwo variable has a sharp upper limit at $m_W$ for  $W$-boson induced backgrounds, while for the $\ttbar + \etmiss$ signal the presence of  extra~\etmiss leads to a tail in the \mttwo distribution. All the other backgrounds, in particular those including the leptonic decays of $Z$ bosons have rapidly falling \etmiss  and \mttwo distributions, and can therefore be strongly suppressed by appropriate cuts on these variables. 

\begin{figure}[t!]
\begin{center}
\includegraphics[width=0.65\textwidth]{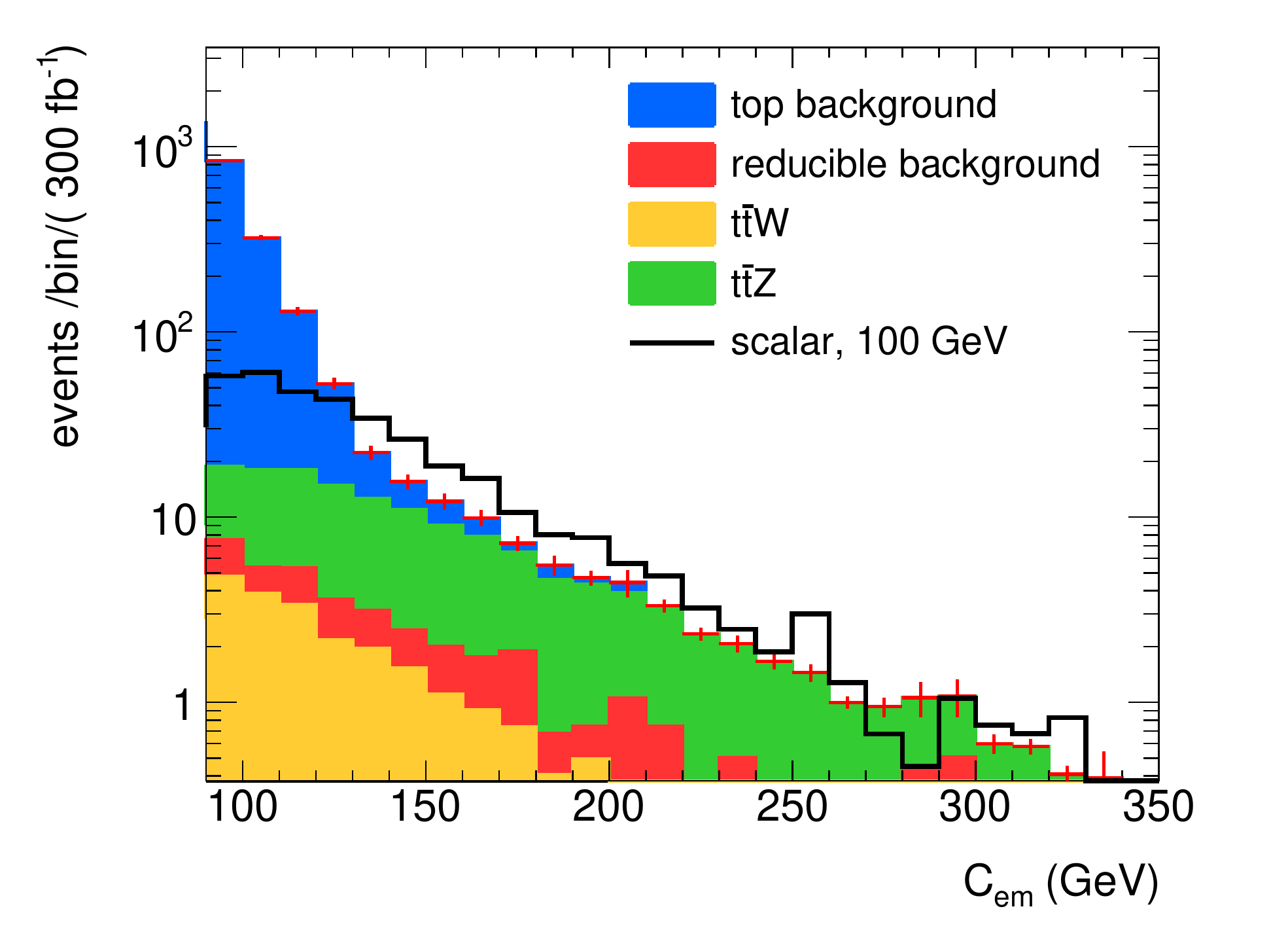}
\caption{Distribution of the $C_{\rm em}$ variable after preselection, basic background suppression and imposing the $\etmiss$ and $\mttwo$ cuts as detailed in the text. The coloured histograms are stacked and represent the different SM backgrounds, while the black line corresponds to the $C_{\rm em}$ distribution as predicted by  a   scalar mediator with $\Msca = 100 \, {\rm GeV}$, $m_\chi = 1 \, {\rm GeV}$ and $g_\chi = g_t = 1$. All predictions have been obtained for $300 \, {\rm fb}^{-1}$ of $14 \, {\rm TeV}$ LHC data and the red error bars are the statistical errors of our background simulations.} 
\label{fig:cem}
\end{center}
\end{figure}

\begin{figure}[t!]
\begin{center}
\includegraphics[width=0.49\textwidth]{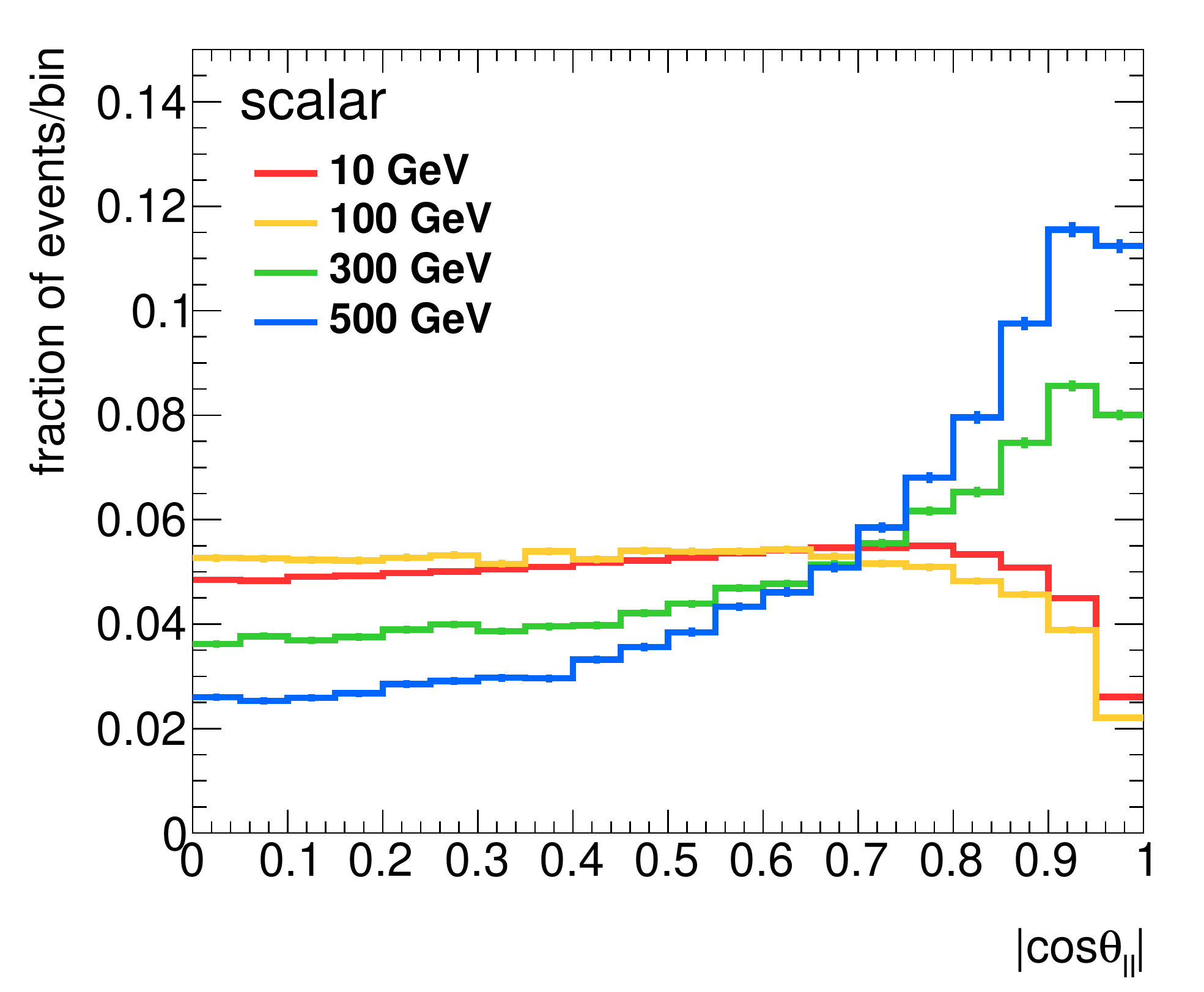}
\includegraphics[width=0.49\textwidth]{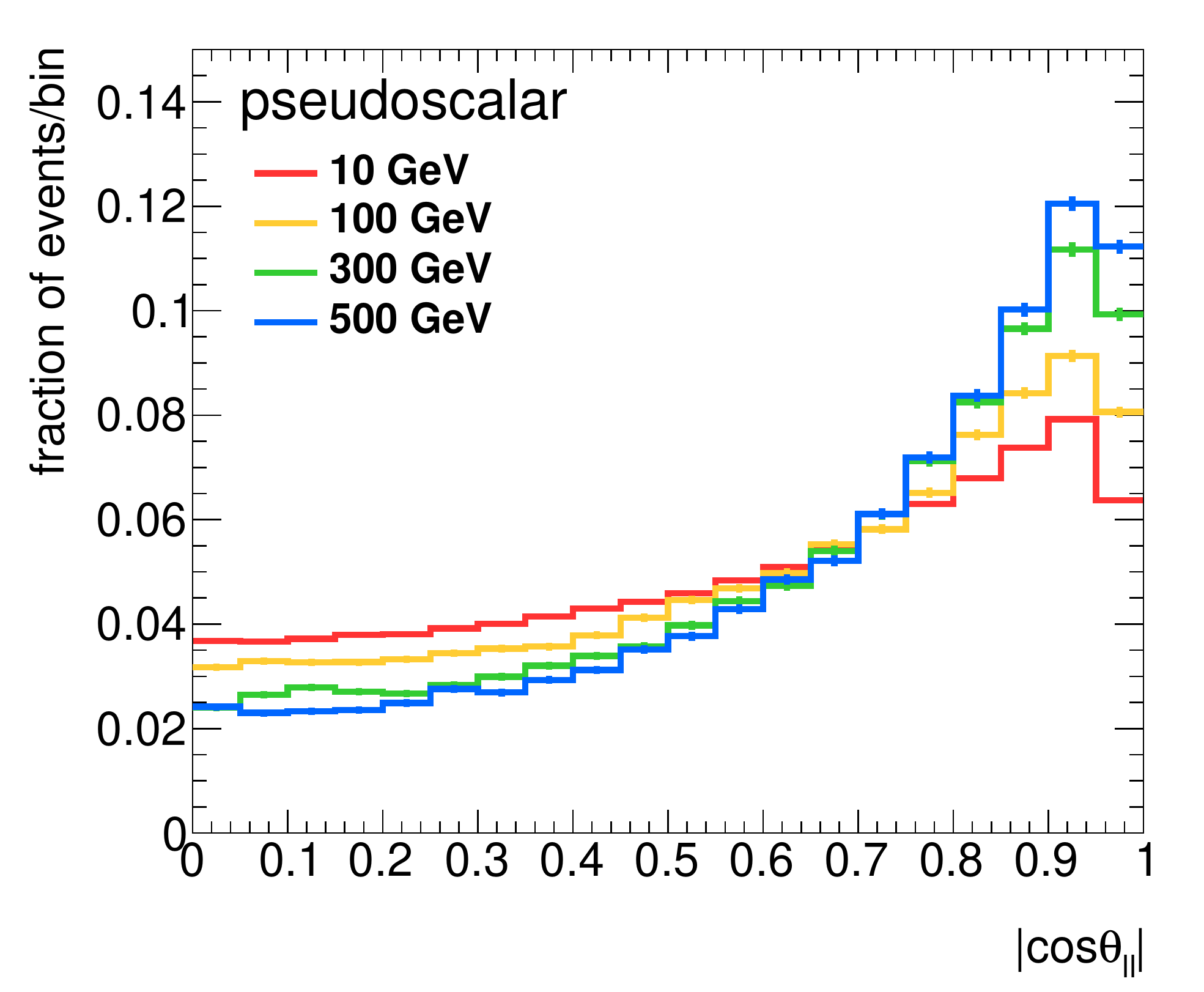}
\includegraphics[width=0.49\textwidth]{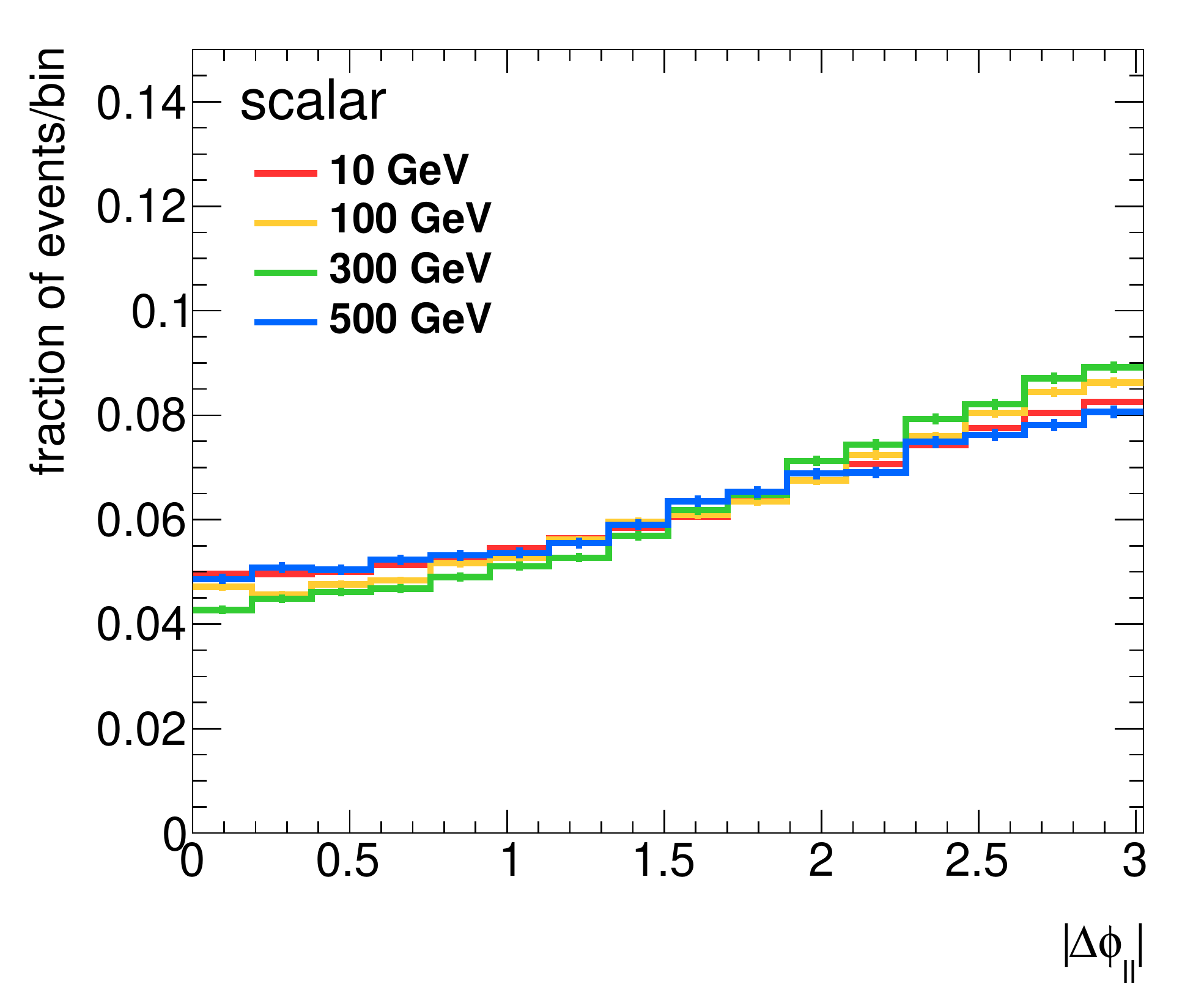}
\includegraphics[width=0.49\textwidth]{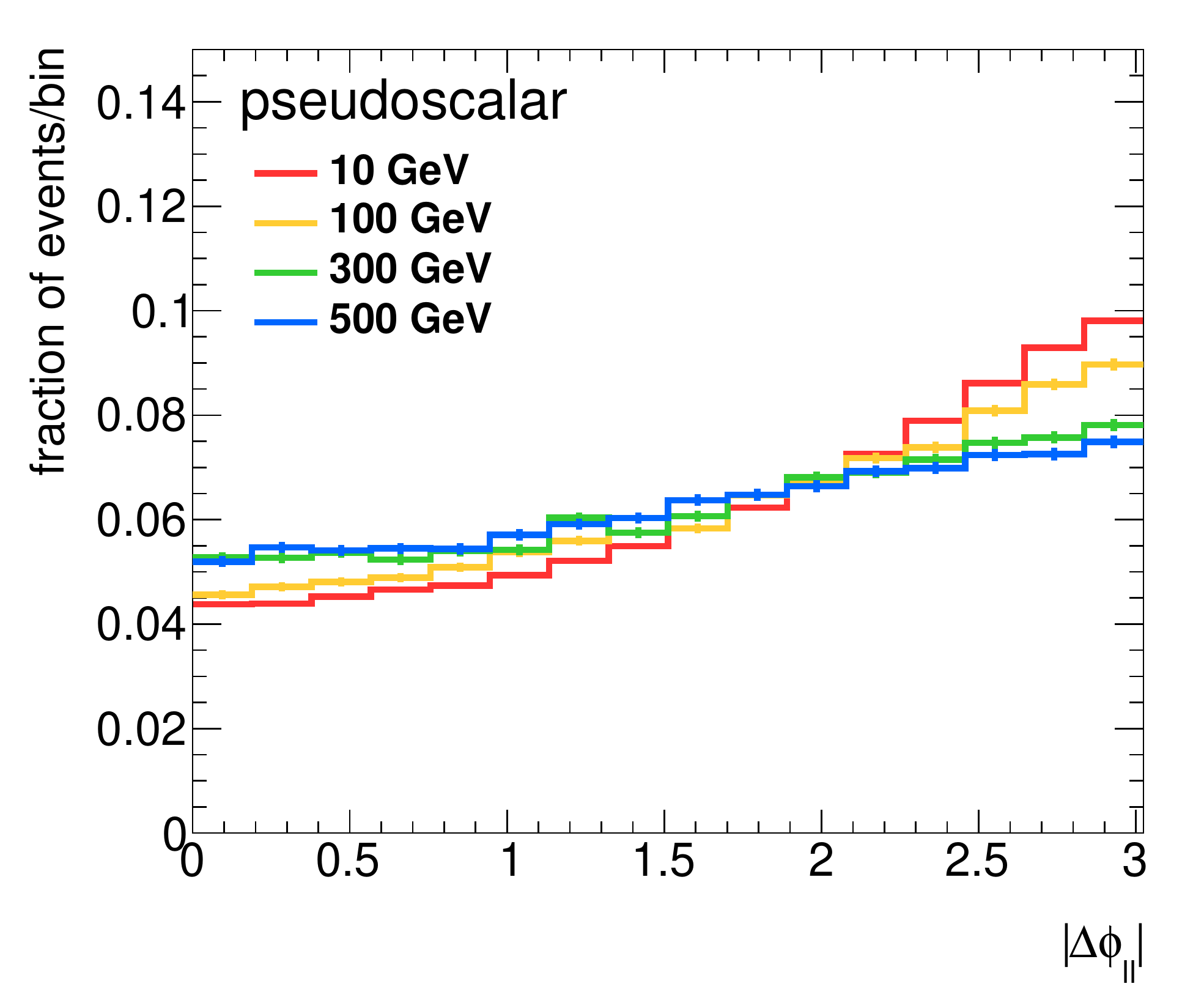}
\caption{Normalised distributions of the $|\hspace{-0.4mm} \cos\theta_{\ell\ell}|$  (upper row) and $|\Delta\phi_{\ell\ell}|$ (lower row)  variables for four different~scalar (left) and  pseudoscalar (right) benchmark models before imposing any selection. The style and colour coding of the curves follows the one of Figure~\ref{fig:ttangular}.  The shown error bars are the statistical errors associated to our MC simulations. }
\label{fig:llfig}
\end{center}
\end{figure}

The first step in the analysis is the preselection. Events pass the preselection only if they have exactly two isolated oppositely charged  leptons (electrons, muons or one of each flavour) with $p_\mathrm{T}^{\ell_1} >25 \, {\rm GeV}$, $p_\mathrm{T}^{\ell_2} >20 \, {\rm GeV}$, $|\eta_{\ell}|<2.5$ and an  invariant mass that satisfies  $m_{\ell \ell} > 20 \, {\rm GeV}$. The $\eta_{\ell}$ and $p_\mathrm{T}^\ell$ requirements ensure that leptons are reconstructed with high efficiency. If the charged signal leptons are of the same flavour the additional requirement $m_{\ell \ell} \in [71, 111] \, {\rm GeV}$ is imposed to veto events where the charged lepton pair  arises from a~$Z \to \ell^+ \ell^-$ decay. 

\begin{figure}[t!]
\begin{center}
\includegraphics[width=0.49\textwidth]{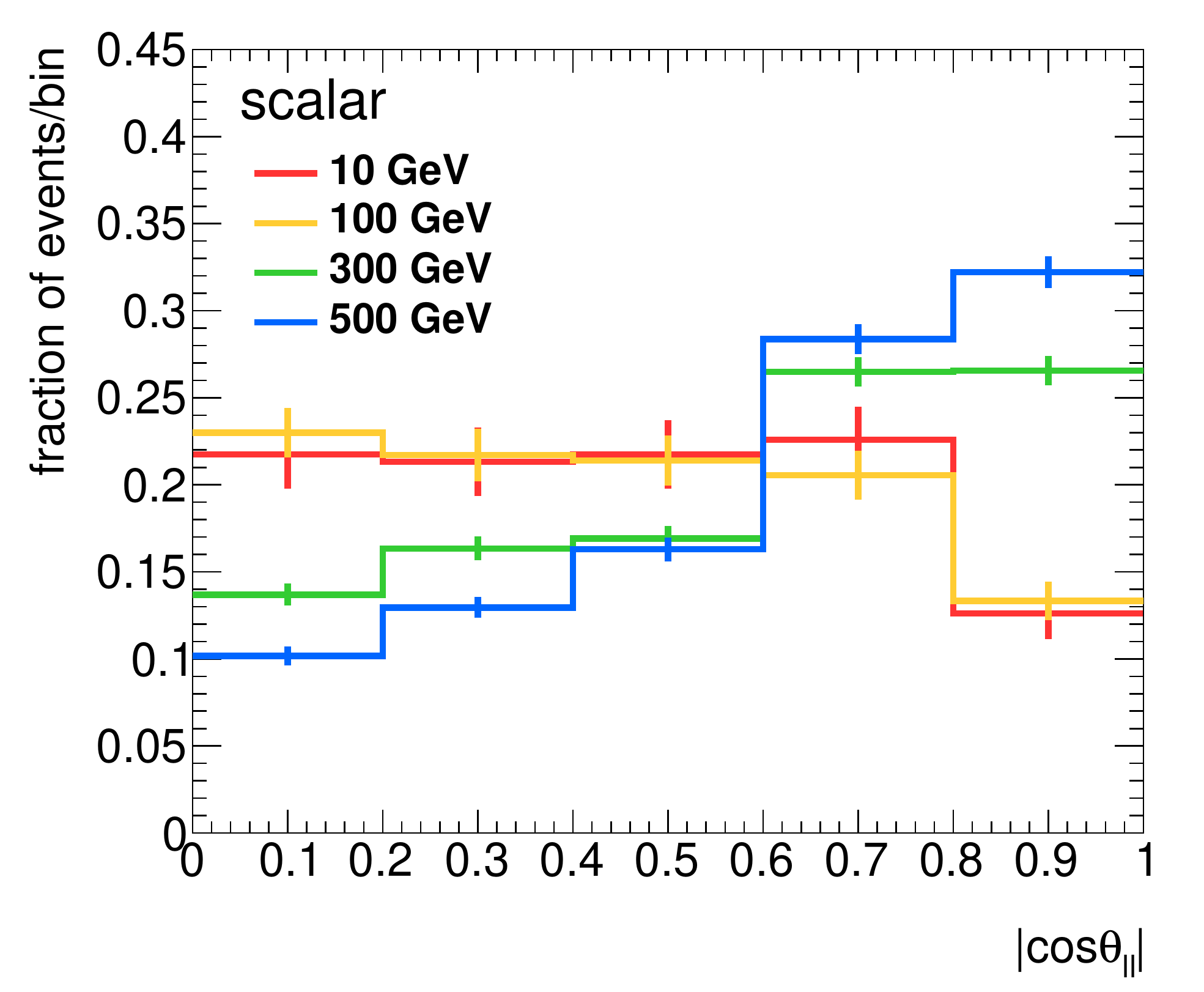}
\includegraphics[width=0.49\textwidth]{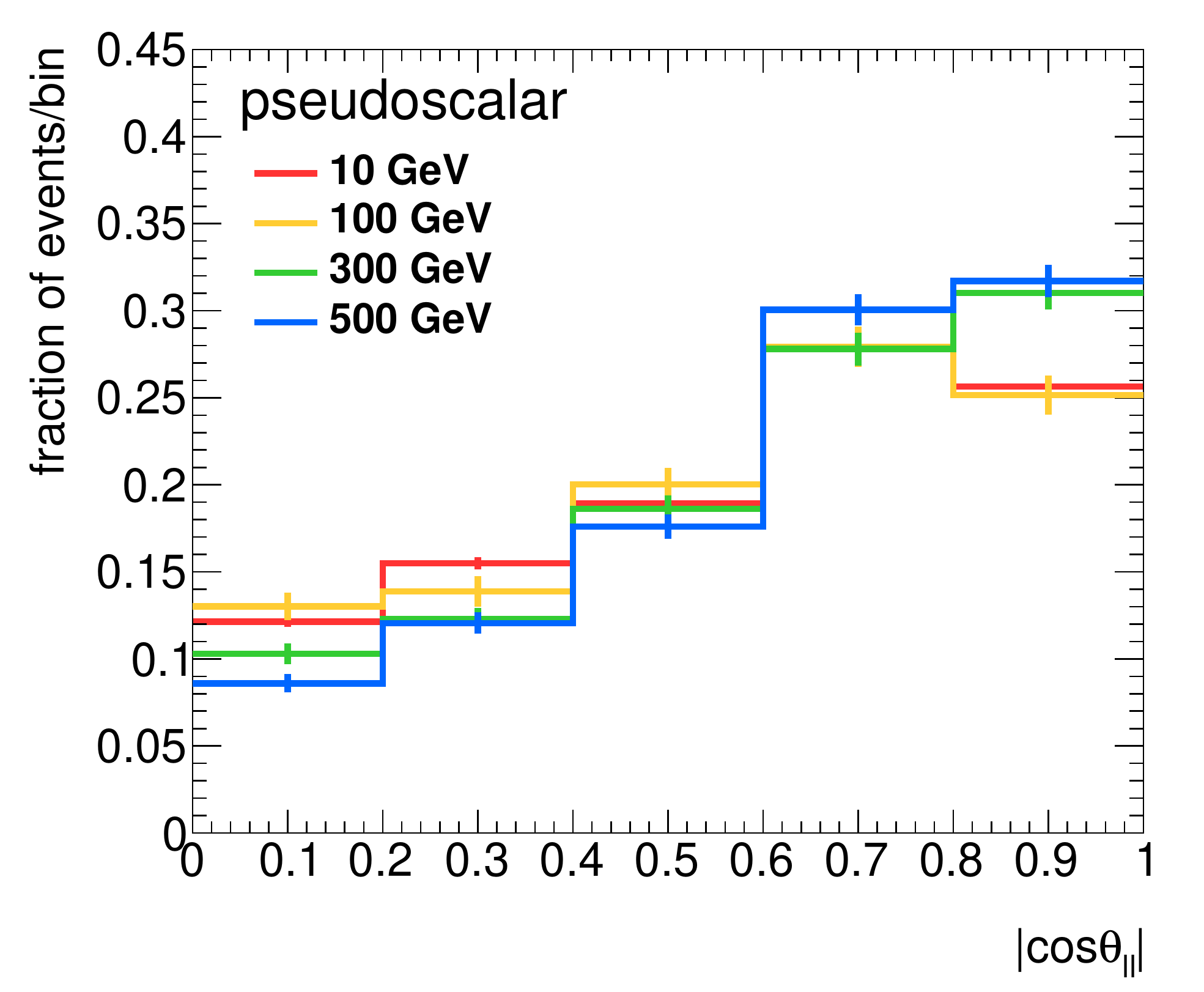}
\includegraphics[width=0.49\textwidth]{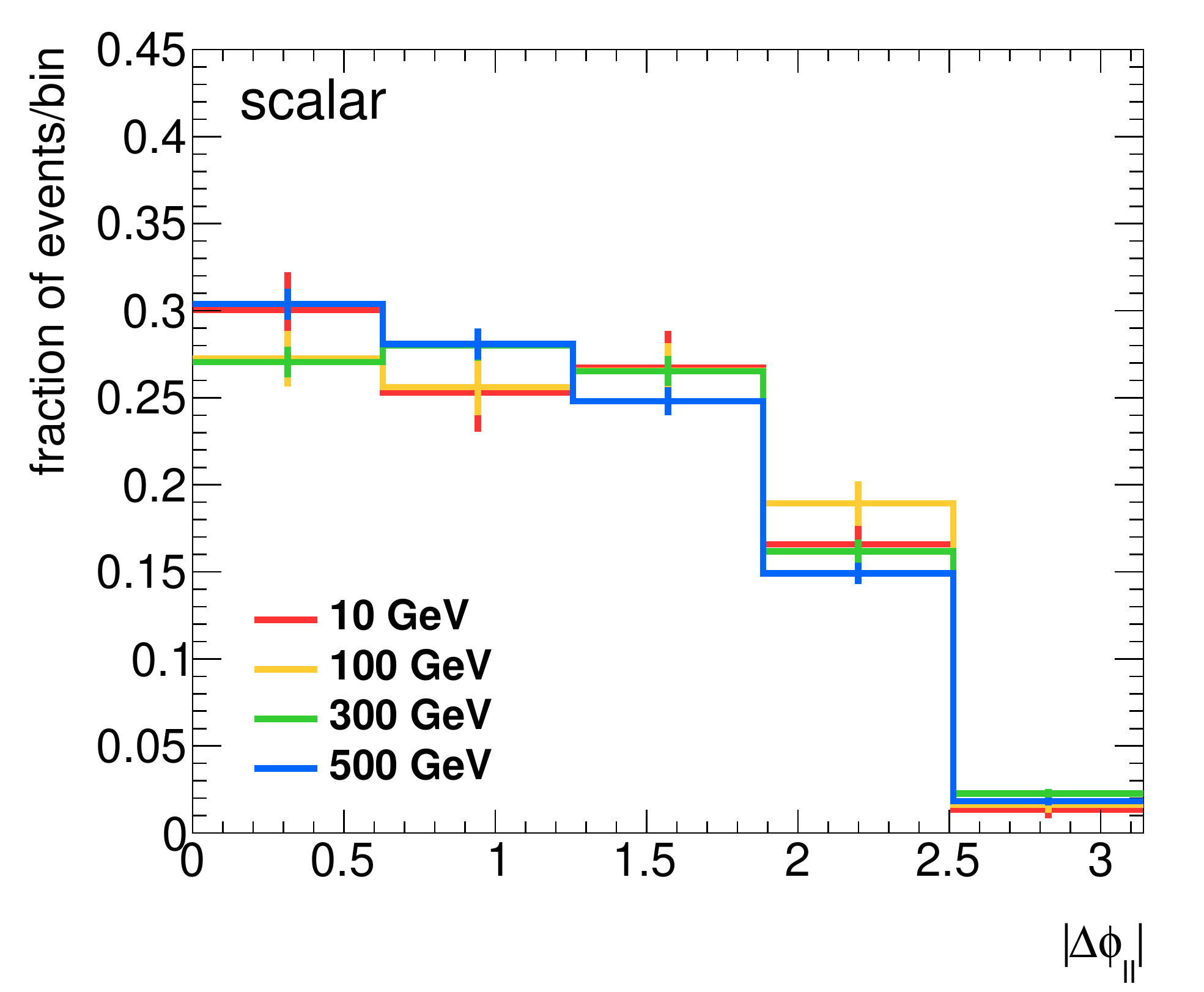}
\includegraphics[width=0.49\textwidth]{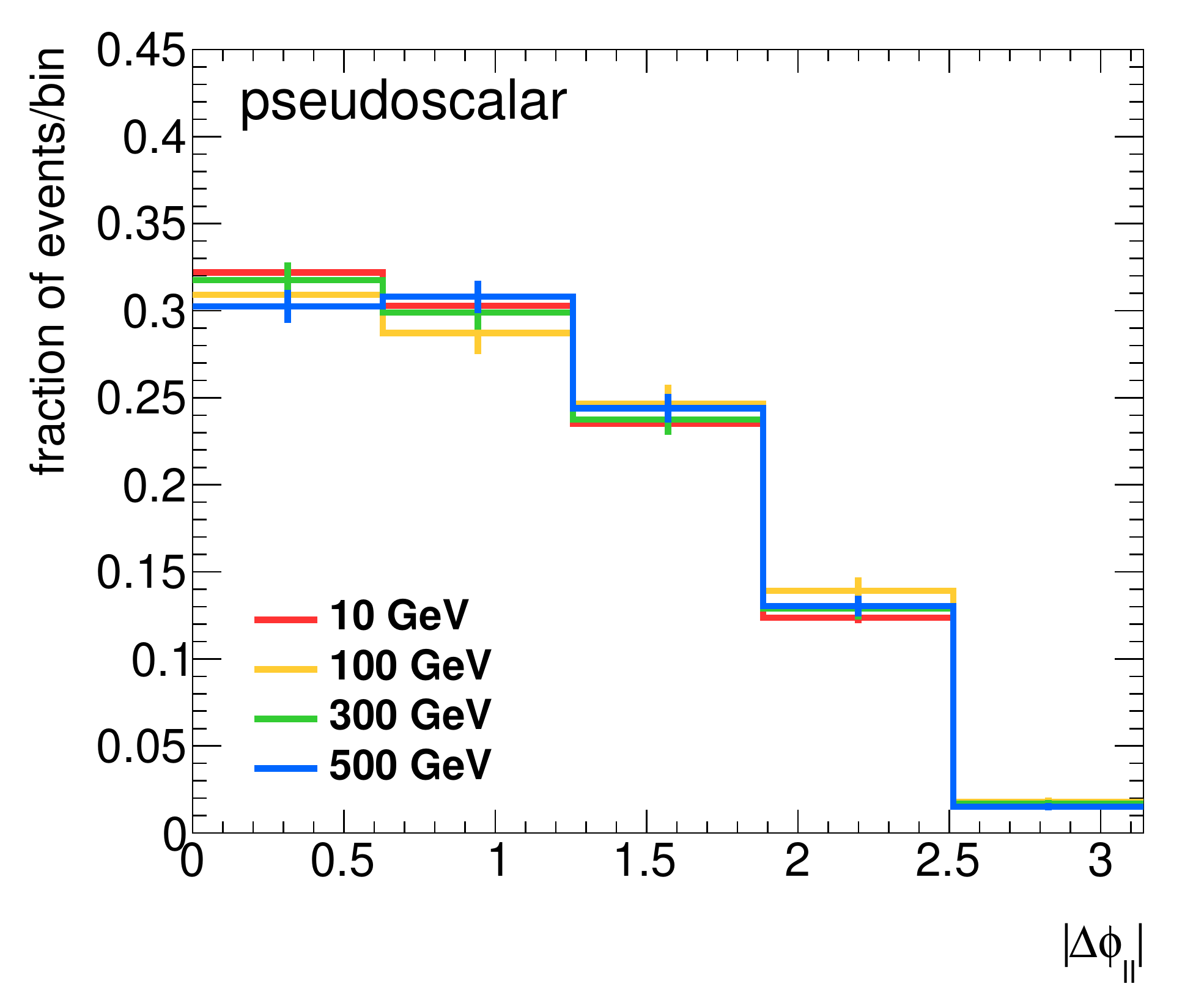}
\caption{Normalised distributions of the $|\hspace{-0.4mm} \cos\theta_{\ell\ell}|$  (upper row) and $|\Delta\phi_{\ell\ell}|$ (lower row)  variables for four different~scalar (left) and  pseudoscalar (right) benchmark models after imposing all selection requirements. The style and colour coding of the curves resembles the one of Figure~\ref{fig:ttangular}. The shown error bars are the statistical errors associated to our MC simulations.}
\label{fig:varcut}
\end{center}
\end{figure}

A further set of cuts aims at a basic reduction of the top and reducible backgrounds. From an inspection of the topology of the top events with both high $\etmiss$ and $\mttwo$, we find that the $\ttbar$ system often recoils against a high-$p_{\rm T}$ jet. For events in  which the leading jet~($j_1$) comes from the decay of a top, the minimum  invariant mass of $j_1$ with the two leptons~$m_{j_1\ell\ell}^{\rm min}$ in the event has to be lower than about $150 \, {\rm GeV}$.  As an  additional requirement we thus impose $m_{j_1\ell \ell}^{\rm min}<150 \, {\rm GeV}$ which rejects events with a high-$p_{\rm T}$  jet produced by QCD radiation, thereby suppressing both the top and reducible backgrounds. The $m_{j_1\ell \ell}^{\rm min}$ cut has an efficiency of around $90\%$ on the signal. In order to further suppress the reducible backgrounds to well below the level of the top backgrounds, the following requirements are employed. Events are required to have at least one $b$-jet with $p_\mathrm{T}^j >30 \, {\rm GeV}$ and all reconstructed jets with $p_\mathrm{T}^j >25 \, {\rm GeV}$ within $|\eta_{\ell}|<2.5$  have to satisfy $\Delta\phi_{\rm min}>0.2$, where~$\Delta\phi_{\rm min}$ is defined to be the angle between $\vec{p}_{\rm T}^{\ j}$ and $\ptmiss$ for the jet closest to $\etmiss$ in the azimuthal plane. The latter requirement suppresses events where the $\etmiss$  is  in part an artefact of jet mismeasurement.

The distribution of events in the  $\etmiss \hspace{0.1mm}$--$\,\mttwo$  plane  after applying the first two sets of cuts is shown in Figure~\ref{fig:etm_mttwo} for the three  classes of SM backgrounds and for a benchmark signal point. The signal prediction corresponds to a scalar mediator with mass $\Msca = 100 \, {\rm GeV}$ and assumes $m_\chi = 1 \, {\rm GeV}$ and $g_\chi = g_t = 1$.  The edge structure of the $\mttwo$ variable for the  $W$-boson induced backgrounds  naturally imposes the requirement $\mttwo>90 \, {\rm GeV}$, whereas a  selection $\etmiss>150 \, {\rm GeV}$ ensures the robustness of the analysis for backgrounds with instrumental $\etmiss$.  From the distributions of events in the upper right panel, one observes that imposing these selection criteria strongly suppresses the reducible backgrounds. To further reduce the top backgrounds, we construct the following linear combination from~$\etmiss$ and~$\mttwo$: 
\beq \label{eq:Cem}
C_{\rm em} \equiv \mttwo + 0.2 \cdot (200 \, {\rm GeV} -\etmiss) \,.
\eeq
The $C_{\rm em}$ distribution after all other selections requirements have been applied is shown in~Figure~\ref{fig:cem} for the various backgrounds and our benchmark signal. The optimal cut on~$C_{\rm em}$ for the benchmark signal was established by minimising the value of the coupling $g = g_\chi = g_t$ which  can be excluded at 95\% CL for an integrated luminosity of 300~fb$^{-1}$, resulting in a requirement $C_{\rm em} > 130 \, {\rm GeV}$. It was checked explicitly that this requirement provides an adequate  sensitivity over the whole considered range of mediator masses.

Figure~\ref{fig:cem} shows that the chosen criteria allows for an adequate reduction of the top backgrounds, while keeping an acceptable signal statistics for the considered model point. After applying the $C_{\rm em}$ cut the residual background is dominated by $\ttbar Z$ with subsequent~$Z\rightarrow\nu_\ell \bar \nu_\ell$ decays. The area defined by the combined $\etmiss$, $\mttwo$ and $C_{\rm em}$ requirements is indicated by the black lines in the panels of  Figure~\ref{fig:etm_mttwo}. One sees that after imposing all three selection criteria the huge majority of events arising from $\ttbar$ and $tW$ are rejected. The total background for an integrated luminosity of $300 \, {\rm fb}^{-1}$ is approximately 100 events of which~$60\%$ are from~$\ttbar Z$, $20\%$ are associated to the top backgrounds, $10\%$ are from reducible backgrounds and $10\%$~are due to~$\ttbar W$. In the mass range from $10 \, {\rm GeV}$ to $500 \, {\rm GeV}$ the signal efficiency varies from $2\mbox{\textperthousand}$ to $2\%$ ($6 \mbox{\textperthousand}$  to $2\%$)  for scalar (pseudoscalar) mediators.

Having developed a realistic strategy for the detection of the DM signal, we can now come back to the angular variables  and study to which extent the $\cos\theta_{\ell\ell}$ and $\Delta\phi_{\ell\ell}$ distributions are distorted by the selection requirements. To illustrate the impact of the selections, we show in~Figures~\ref{fig:llfig} and~\ref{fig:varcut} predictions for the  angular distributions for eight different benchmark models before and after imposing the full set of selection cuts. From the panels in~Figure~\ref{fig:llfig} one observes that if no cuts are imposed both $|\hspace{-0.4mm} \cos\theta_{\ell\ell}|$ and~$|\Delta\phi_{\ell\ell}|$ provide sensitivity to the mass and the CP nature of the mediator particle. Even before applying any event selection the shape differences observed in $|\hspace{-0.4mm} \cos\theta_{\ell\ell}|$ are however more pronounced than those in $|\Delta\phi_{\ell\ell}|$. 

The results presented in Figure~\ref{fig:varcut} demonstrate that  applying the signal selections has a notable impact on the distributions of both angular observables. However, while for the $|\hspace{-0.4mm} \cos\theta_{\ell\ell}|$ variable the shape of the distributions still differ visibly for the chosen mediator masses and  types, in the case  of $|\Delta\phi_{\ell\ell}|$ almost all  differences are washed out once experimental selections cuts are employed. Notice that the distortion of the $|\Delta\phi_{\ell\ell}|$ spectra is most pronounced for values of $|\Delta\phi_{\ell\ell}|$ close to $\pi$, which corresponds to kinematic configurations where the two leptons are back-to-back in the azimuthal plane. The imposed~$\mttwo$ cut typically removes such events thereby strongly reducing the discriminating power of the azimuthal angle difference $\Delta\phi_{\ell\ell}$.\footnote{The recent analysis \cite{Buckley:2015ctj} that uses $\Delta\phi_{\ell\ell}$ as a CP analyser claims that only a cut of $\etmiss > 200 \, {\rm GeV}$ but no $\mttwo$ selection is needed to reduce the SM backgrounds to manageable levels. We are unable to reproduce this finding.}  The observable~$|\hspace{-0.4mm} \cos\theta_{\ell\ell}|$ instead still shows promising discriminating properties in a realistic experimental analysis with  $\etmiss$ and $\mttwo$ cuts. We will therefore use this observables in the next section as our CP analyser. 

\section{Results}
\label{sec:results}

\begin{figure}[t!]
\begin{center}
\includegraphics[width=0.49\textwidth]{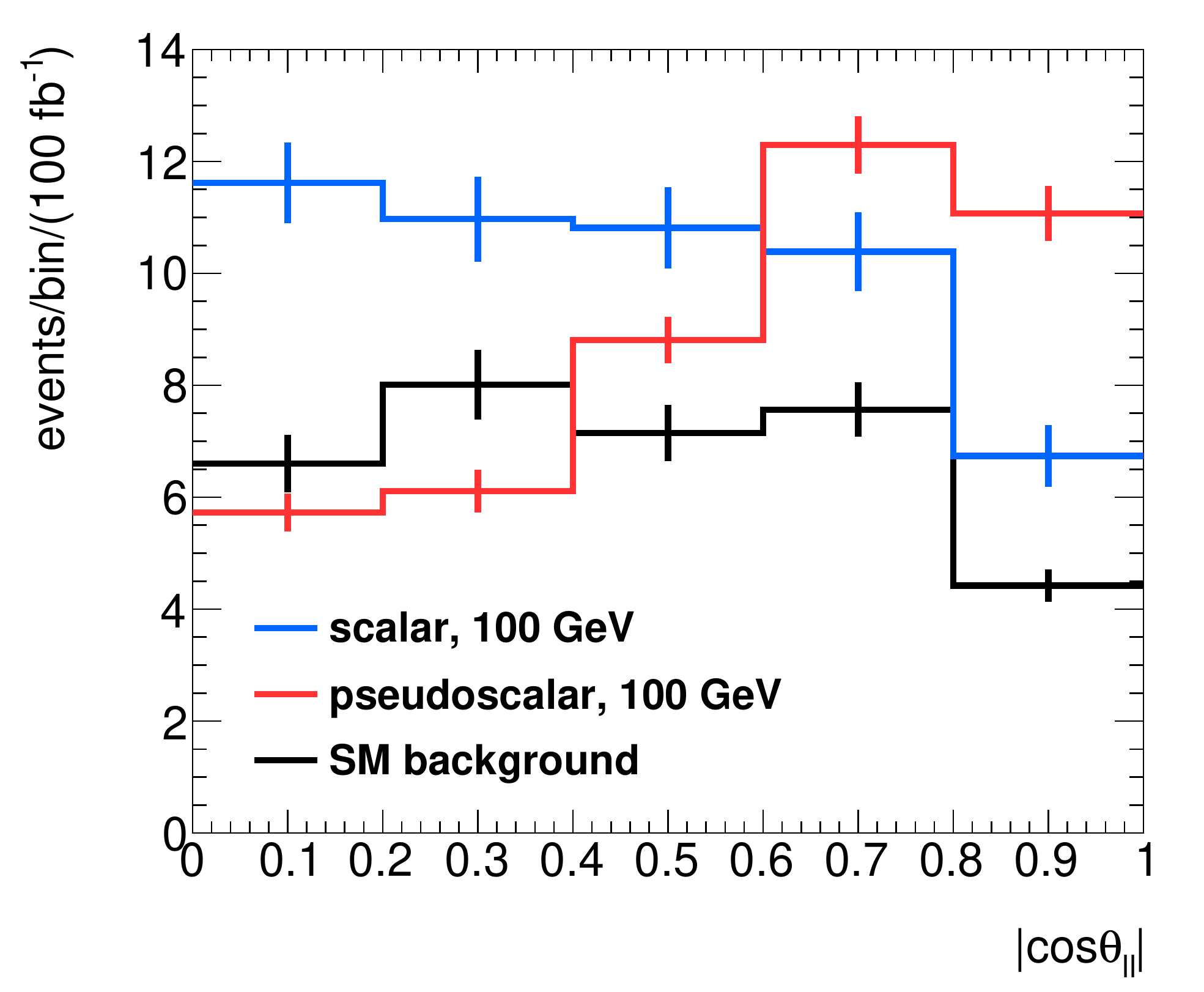}
\caption{Distribution of the $|\!\cos\theta_{\ell\ell}|$ variable after employing the full selection requirements as specified in Section~\ref{sec:analysis}. The normalisation corresponds to the numbers of events expected for $100 \, {\rm fb} ^{-1}$ at $\sqrt{s} = 14 \, {\rm TeV}$. The error bars indicate the errors on the generated MC statistics.}
\label{fig:costhbg}
\end{center}
\end{figure}

On the basis of the selection criteria defined in the previous section, we can study the LHC sensitivity to the $\ttbar + \etmiss$ signature  for an integrated  luminosity of $300 \, {\rm fb}^{-1}$ and $3 \, {\rm ab}^{-1}$ at $\sqrt{s} = 14 \, {\rm TeV}$. A profiled likelihood ratio test statistic is used to evaluate the upper limit on the ratio of the signal yield to that predicted in the DMF spin-0 simplified models with $m_\chi = 1 \, {\rm GeV}$ and $g_\chi = g_t =1$. This ratio will be called signal strength in the following and denoted by~$\mu$. The $CL_s$ method \cite{Read:2002hq} is used to derive the confidence level~(CL) of the exclusion limits and signal models with $CL_s$ values below $0.05$ are said to be excluded at~95\%~CL.
The statistical analysis has been performed employing the {\tt HistFitter} toolkit~\cite{Baak:2014wma}.

Our sensitivity study is performed in two ways. First by performing a simple counting experiment and second by including shape information in the form of a  5-bin likelihood fit to the $|\!\cos\theta_{\ell\ell}|$ distributions. The inclusion of shape information is motivated by the observation that the distributions of events as a function of the pseudorapidity difference of the  dilepton pair is different for signal and background. This feature is illustrated in Figure~\ref{fig:costhbg} which compares the predictions for a scalar (blue curve) and pseudoscalar (red curve) assuming $M = 100 \, {\rm GeV}$, $m_\chi = 1 \, {\rm GeV}$ and $g_\chi = g_t = 1$ with the SM background~(black~curve).  

Given the presence of a sizeable irreducible background surviving all the selections, the experimental sensitivity will be largely determined by the systematic uncertainty on the estimate of the SM backgrounds. Such an error has  two main sources: on the one hand, uncertainties on the parameters of the detector performance such as the energy scale for hadronic jets and the identification efficiency for leptons, and on the other hand, uncertainties plaguing the MC modelling of SM processes.  Depending on the process and on the kinematic selection, the total uncertainty can vary between a few percent and a few tens of percent. The present analysis does not select extreme kinematic configurations for the dominant $\ttbar Z$ background, and it therefore should be possible to control the experimental systematics at the 10\% to 30\% level.  In the following, we will assume a systematic error of~20\% on the backgrounds in the case of the counting experiment. In the case of the~5-bin shape fits we will consider background uncertainties of both 30\% and 20\%, fully correlated across the bins. We have checked that in the absence of an external measurement (e.g.~a background control region) which profiles uncertainties, the use of correlated uncertainties in the shape fit provides the most conservative results.

\begin{figure}[t!]
\begin{center}
\includegraphics[width=0.49\textwidth]{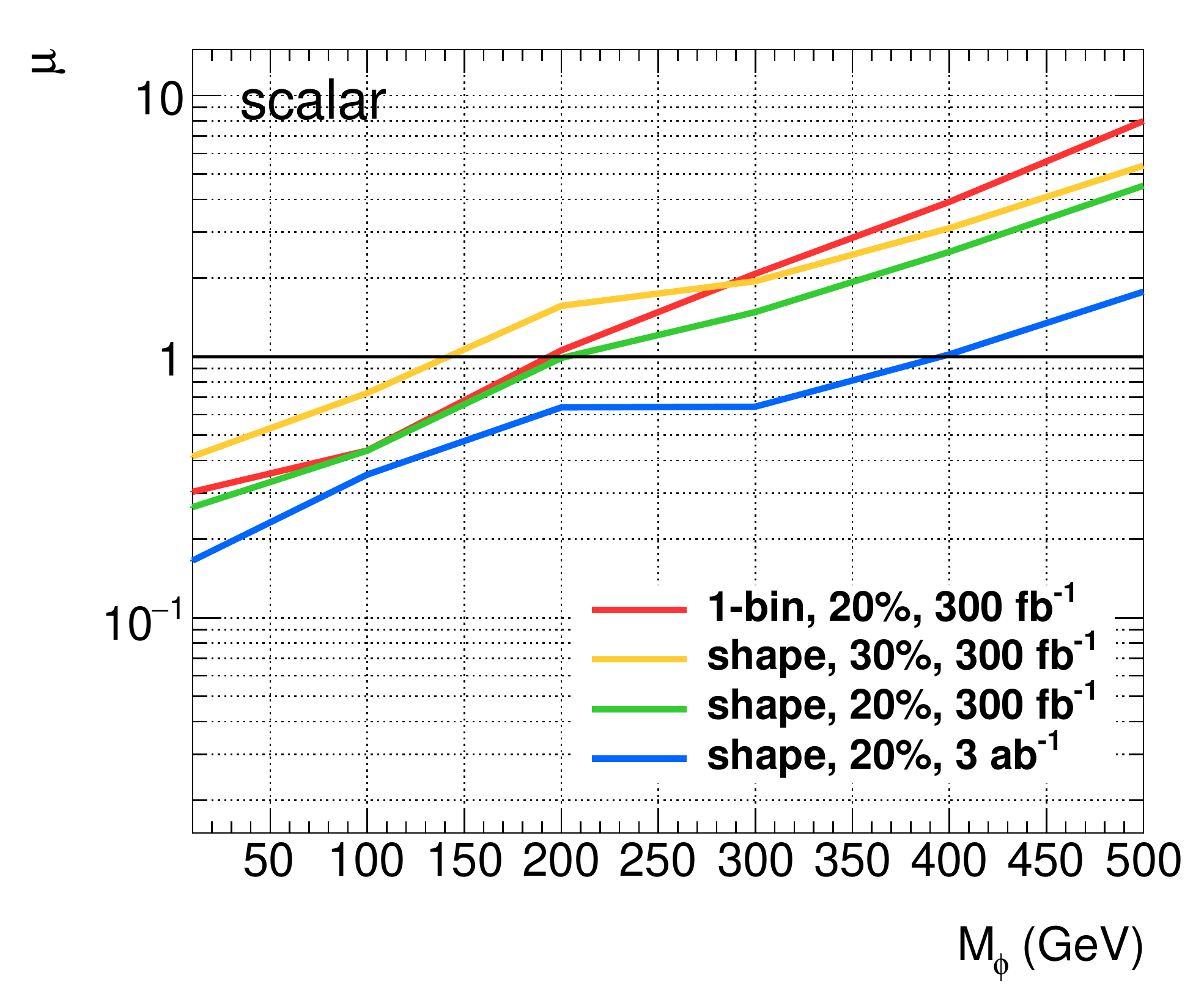}
\includegraphics[width=0.49\textwidth]{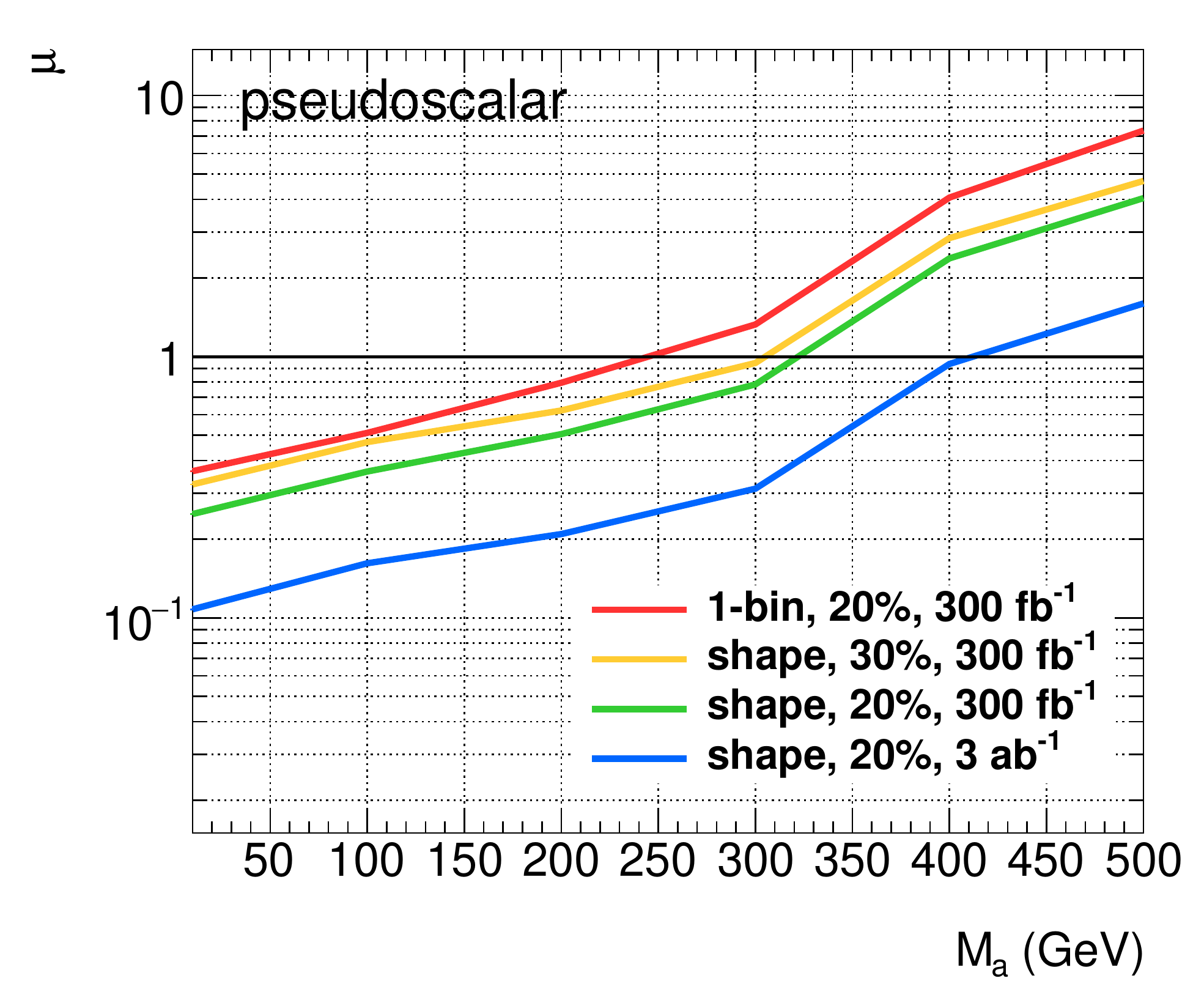}
\caption{Value of the signal strength that can be excluded at 95\% CL as a function of  the mass for scalar (left) and pseudoscalar (right) mediators. The reach with $300 \, {\rm fb}^{-1}$ of $\sqrt{s} = 14 \, {\rm TeV}$ data is given for a simple counting experiment assuming a 20\% systematic background uncertainty~(red curves) and for 5-bin shape fits with both 30\%~(yellow curves) and 20\%~(green curves) errors. A~hypothetical shape-fit scenario based on  $3 \, {\rm ab}^{-1}$ and 20\% systematics is also shown (blue curves).}
\label{fig:95cl}
\end{center}
\end{figure}

The results of our sensitivity study are displayed in Figure~\ref{fig:95cl}.  Notice that the results shown for $3 \, {\rm ab}^{-1}$ rely on the assumption that the $\etmiss$ measurement performance in the very harsh experimental conditions of the HL-LHC will be equivalent to the one achieved during LHC~Run~I. As expected from the shapes of the distributions  in Figures~\ref{fig:varcut} and~\ref{fig:costhbg}, the 5-bin likelihood fit provides a significant improvement over the counting experiment for high-mass mediators irrespectively of their~CP nature. The gain in sensitivity at lower mass depends  strongly on the assumption on the systematic uncertainty of the SM background. For instance assuming a~20\% systematics on the counting experiment and a~30\% background error on the shape fit, we find that the shape analysis will have larger discriminating power than the simple cut-and-count strategy for $\Msca \gtrsim 300 \, {\rm GeV}$ and $\Mpse \gtrsim 100 \, {\rm GeV}$ with $300 \, {\rm fb}^{-1}$ of integrated luminosity.  If the background for the shape fit can instead be estimated with an error of~20\%, including shape information is expected to be the superior strategy over almost the entire range of considered masses. In fact, at the LHC with $3 \, {\rm ab}^{-1}$ of data it should be possible to exclude spin-0 models that predict $\mu =1$ for mediator masses up to around $400 \, {\rm GeV}$ using the~5-bin likelihood fit employed in our study. The observed strong dependence of the reach on the assumption on the systematic background uncertainty shows that  a good  experimental understanding of $\ttbar Z$ production within the SM will be a key ingredient to  a possible discovery of DM in the $\ttbar + \etmiss$ channel. 

\begin{figure}[t!]
\begin{center}
\includegraphics[width=0.49\textwidth]{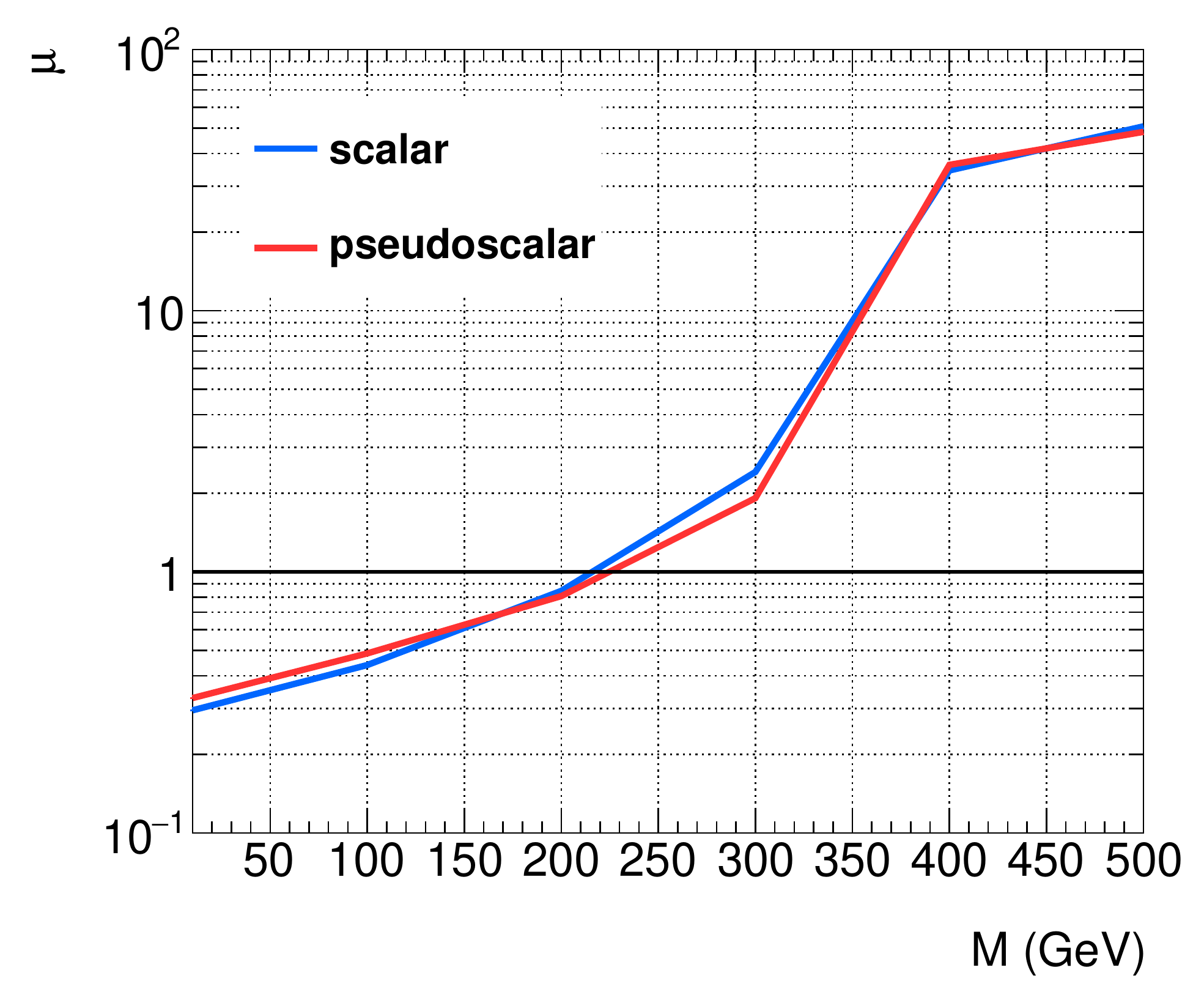}
\caption{Value of the signal strength for which the scalar (pseudoscalar) hypothesis can be excluded at 95\%~CL in favour of the pseudoscalar (scalar) hypothesis. The reach is given for  a shape fit using the $|\!\cos\theta_{\ell\ell}|$ variable, assuming $300 \, {\rm fb}^{-1}$ of $14 \, {\rm TeV}$ LHC data and a systematic  uncertainty of 20\% on the SM background.}
\label{fig:95clSpin}
\end{center}
\end{figure}

We also perform a hypothesis test between the scalar and pseudoscalar mediator hypotheses as a function of the mediator mass. Figure~\ref{fig:95clSpin} shows the value of $\mu$  for which the scalar hypothesis can be excluded at 95\%~CL in favour of  the pseudoscalar one (blue curve) and vice versa (red curve). Our statistical analysis is based on a~5-bin shape fit of the~$|\!\cos\theta_{\ell\ell}|$ distributions and employs standard maximum likelihood estimator techniques~(see for instance \cite{ATL-PHYS-PUB-2011-11}) that are implemented in the {\tt RooFit/RooStat} package \cite{Verkerke:2003ir}. From the  figure it is evident that based on $300 \, {\rm fb}^{-1}$ of $\sqrt{s} = 14 \, {\rm TeV}$ data and under the assumption that the SM backgrounds can be determined with an uncertainty of 20\%, it should be possible to distinguish between the two CP hypotheses for masses $M \lesssim 200 \, {\rm GeV}$ in scenarios that lead to a signal strength of 1 or lower. 

While we have assumed in our hypothesis test  that $M$ is already known, constraining the mediator mass, its couplings and CP nature simultaneously using our proposal seems feasible. For instance constraining both $M$ and~$g_t$ should be possible by combining the information on the $C_{\rm em}$   variable (or $\mttwo$ observable) with that on the fiducial $\ttbar + \etmiss$ cross section. The obtained constraints may be  further strengthened  by including besides the $\ttbar + \etmiss$ channel other $\etmiss$ measurements such as mono-jets into a global fit.  Such a fit would depend on the detailed consideration of additional search channels and of their experimental uncertainties, and is outside the the scope of the present study.

\begin{figure}[t!]
\begin{center}
\includegraphics[width=0.49\textwidth]{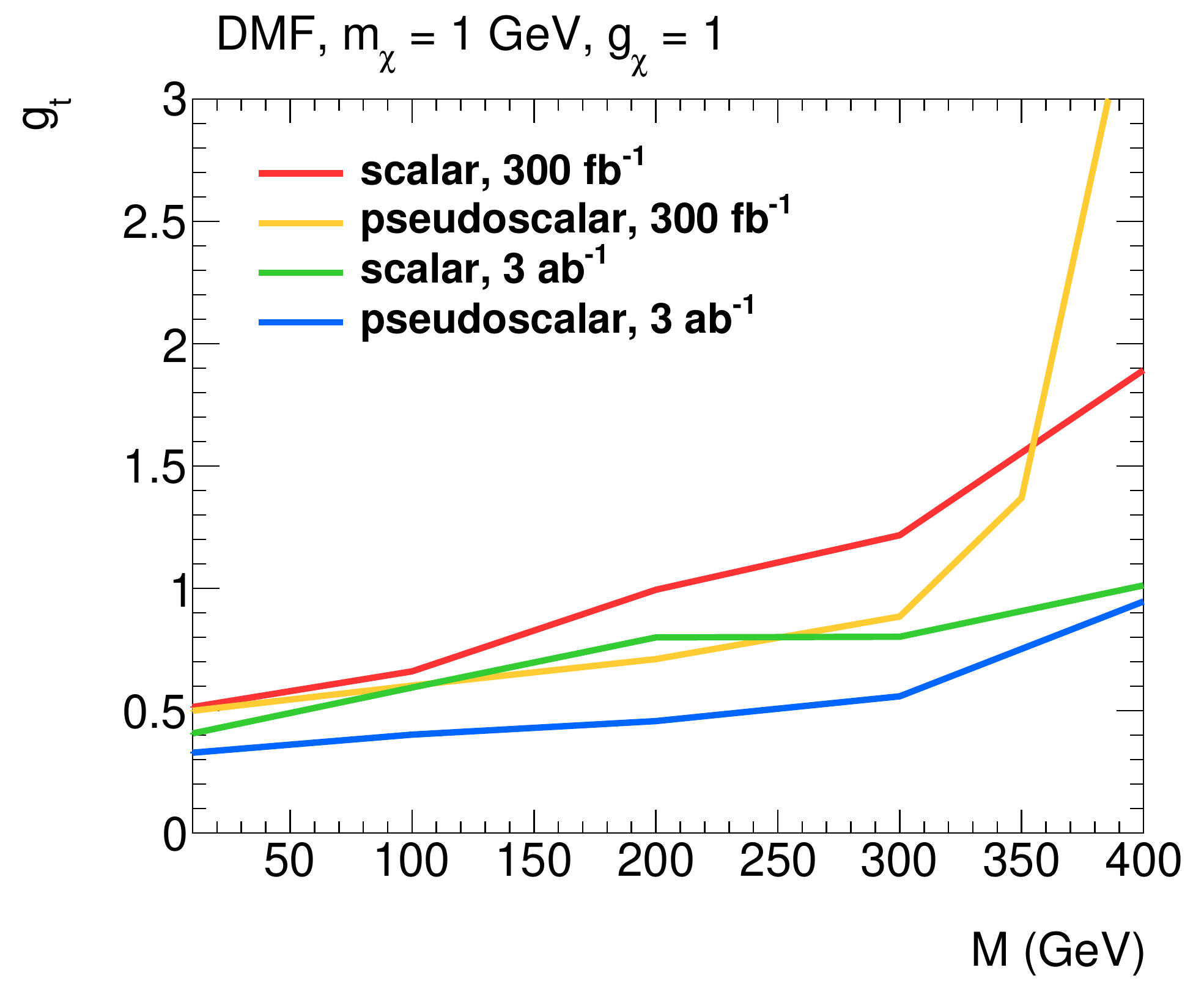}
\includegraphics[width=0.49\textwidth]{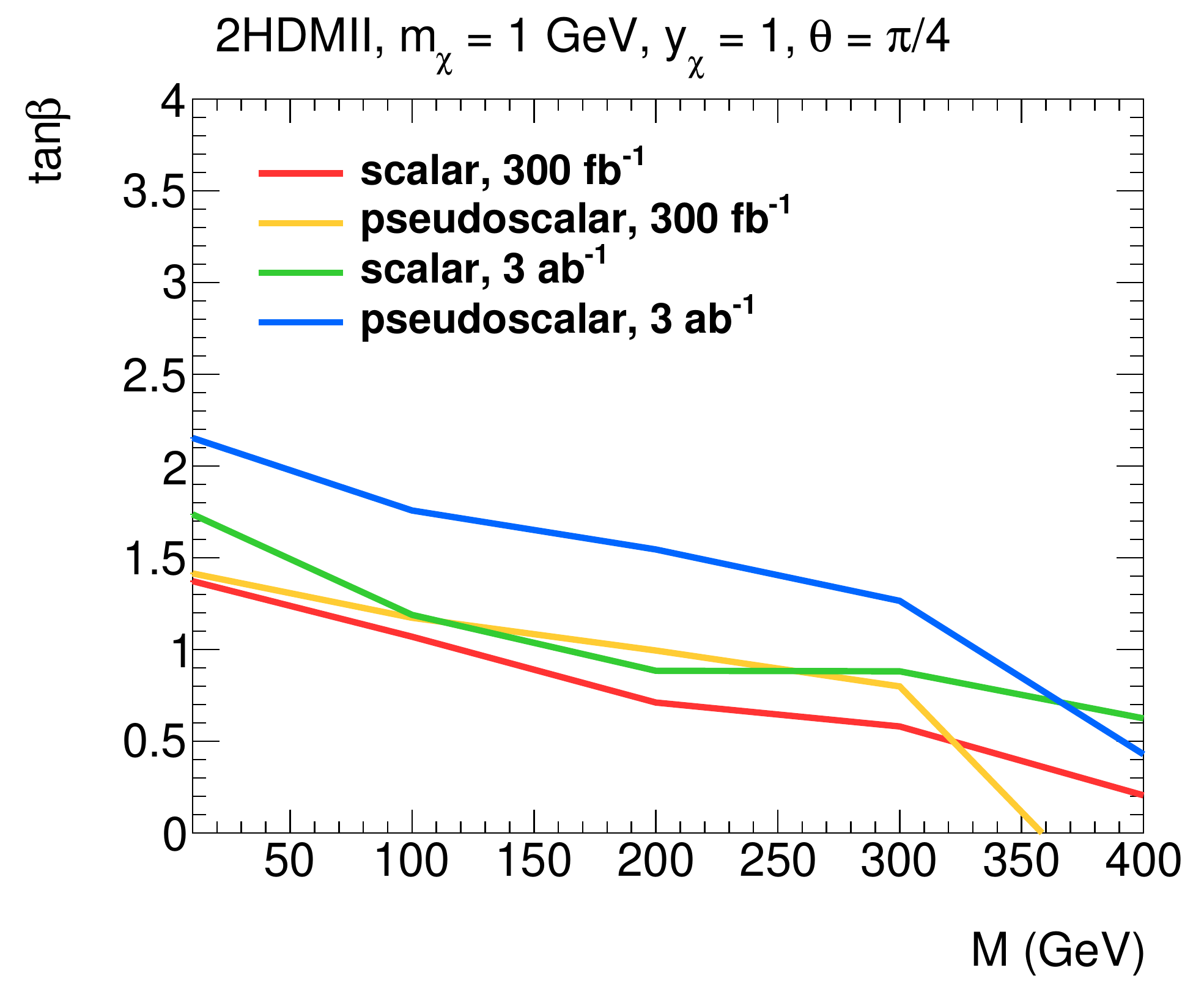}
\caption{Left: Value of the coupling $g_t$ that can be excluded at 95\%~CL in the DMF spin-0 models. The shown limits correspond to $m_\chi = 1 \, {\rm GeV}$ and $g_\chi = 1$ and the parameter space above the coloured curves is ruled out. Right: Value of $\tan \beta$ that can be excluded at 95\%~CL in the alignment/decoupling limit of the 2HDMII plus singlet models.  The relevant model parameters are $m_\chi = 1 \, {\rm GeV}$, $y_\chi = 1$ and $\theta = \pi/4$ and the exclusion holds for the parameter space to the bottom-left of the coloured curves. All limits have been obtained from our 5-bin shape-fit analysis assuming  a systematic error of 20\% on the SM background. }
\label{fig:interpretations}
\end{center}
\end{figure}

The model-independent limits on the signal strengths derived in Figure~\ref{fig:95cl} can be used to constrain the parameter space of the two classes of simplified DM models that have been introduced in Section~\ref{sec:simplified}. In the left (right) panel of Figure~\ref{fig:interpretations} we present the 95\%~CL exclusion bounds in the $M$--$\hspace{0.25mm} g_t$ ($M$--$\hspace{0.5mm} \tan \beta$) plane that apply in the case of the DMF (2HDMII plus singlet) models. The constraints are obtained from a shape fit with 20\% background uncertainties and the  red, yellow, green and blue lines illustrate the scalar and pseudoscalar case assuming integrated luminosities of $300 \, {\rm fb}^{-1}$ and $3\, {\rm ab}^{-1}$. From the left panel we see that for the parameter choices $m_\chi = 1 \, {\rm GeV}$ and $g_\chi = 1$ values of $g_t \lesssim 1$ can be ruled out in the mass ranges $\Msca \lesssim 200 \, {\rm GeV}$ and $\Mpse \lesssim 320\, {\rm GeV}$ after $300 \, {\rm fb}^{-1}$ of data have been collected. With ten times more luminosity these exclusions then extend up to mediator masses close to~$400 \, {\rm GeV}$ in the DMF models. 

In the case of the 2HDMII plus singlet models in the alignment/decoupling limit, one sees from the right panel in Figure~\ref{fig:interpretations}  that with  $300 \, {\rm fb}^{-1}$ data model realisations with $\tan \beta \gtrsim 1$ can be excluded for $\Msca \lesssim 120 \, {\rm GeV}$ and $\Mpse \lesssim 200 \, {\rm GeV}$. The corresponding choice of parameters is $m_\chi = 1 \, {\rm GeV}$, $y_\chi = 1$ and $\theta = \pi/4$. With $3 \, {\rm ab}^{-1}$ of integrated luminosity the quoted mass limits are expected to improve to $\Msca \lesssim 160 \, {\rm GeV}$ and $\Mpse \lesssim 330 \, {\rm GeV}$. 

It is also apparent from both panels that below the $\ttbar$ threshold the limits on pseudoscalar models are always stronger than those on scalar scenarios. This feature can be understood by realising that our $|\!\cos \theta_{\ell \ell}|$ shape analysis has larger discriminating power for a CP-odd than for a CP-even spin-0 portal state, as one would expect from Figure~\ref{fig:costhbg}. Notice finally that above the $\ttbar$ threshold the constraints in the   $M$--$\hspace{0.25mm} g_t$  and $M$--$\hspace{0.5mm} \tan \beta$ planes as depicted in  Figure~\ref{fig:interpretations} start to weaken because the branching ratios of $\phi/a \to \chi \bar \chi$ are no longer $100 \%$. This feature is most pronounced in the case of our pseudoscalar scenario with $300 \, {\rm fb}^{-1}$ of data. In this parameter space region ditop resonance searches can provide relevant constraints~\cite{ATLAS-CONF-2016-073,HaischTeVPA16} on both the DMF as well as the 2HDMII plus singlet models. 

\section{Conclusions}
\label{sec:conclusions}

In this article, we have studied the prospects of future LHC runs to probe spin-0 interactions between DM and top quarks via the $\ttbar + \etmiss$ signature. This final state is particularly interesting, since it is expected to have an appreciable  rate in simplified  $s$-channel  scalar and pseudoscalar models that satisfy both the constraints from quark-flavour and Higgs physics. Examples of such models are provided by the spin-0 scenarios recommended by the ATLAS/CMS DMF and 2HDMII plus singlet extensions in the alignment/decoupling limit and low values of $\tan \beta$.

In order to understand which kinematic variables are useful to separate  signal and  SM backgrounds, we have first analysed the basic properties of the $\ttbar + \etmiss$ signal. By identifying the dominant production topologies, qualitative explanations of the mediator mass and type dependence of the total production cross sections~$\sigma$ and their  gluon-fusion fractions~$\sigma_{gg}/\sigma$ have been provided. It has also been shown that by dividing the signal into mediator-fragmentation and top-fusion diagrams allows one to understand the  basic features of the $\cos \theta_{\ttbar}$  distributions   and the azimuthal angle difference~$\Delta \phi_{\ttbar}$ of the $\ttbar$ system. In all cases we have confirmed our general expectations with explicit MC simulations, finding that the angular correlations of the top-antitop pair provide in principle powerful probes of the CP properties of the mediation mechanism in~$\ttbar + \etmiss$ production.

Given the presence of four invisible particles in the dilepton final state, the relative orientation of the  top quarks is however experimentally not directly accessible. Any information on the $\ttbar$ system must thus be indirectly obtained from the kinematical distributions of the dilepton pair resulting from $t \to bW \, (W \to \ell \nu_\ell)$. In our work we have developed a realistic analysis strategy to enhance the small signal-to-background ratio of the $\ttbar + \etmiss$  dilepton channel.  We found that only a combination of selections based on the $\etmiss$ and the~$\mttwo$ variables is able to suppress the overwhelming $\ttbar$ and $tW$ backgrounds sufficiently. The impact of these cuts on  the $\cos \theta_{\ell \ell}$ and $\Delta \phi_{\ell \ell}$ distributions has then been studied in detail. While the pseudorapidity difference of the dilepton pair was found to be affected by the selections in the transverse plane  only in a minor way, the azimuthal angle difference was shown to loose most of its  discriminating power due to the $\mttwo$ requirement. In a realistic experimental search for a $\ttbar + \etmiss$ signal,  the $\Delta \phi_{\ell \ell}$  variable thus seems to be of limited use in probing the  CP nature of the DM $\ttbar$  interactions.

Focusing on the $\cos \theta_{\ell \ell}$ observable we have then presented  a comprehensive sensitivity study for the considered DM  signal.  We have derived the 95\%~CL exclusion limits on the signal strengths $\mu$ as a function of the mediator mass $M$ that follow from both a counting experiment and a $5$-bin shape fit to the $|\! \cos \theta_{\ell \ell}|$ distributions. Our analysis shows that including shape information generically allows to improve the reach of the $\ttbar + \etmiss$ searches. The actual improvement however  turns out to depend sensitively on the assumption about the systematic uncertainty on the irreducible SM backgrounds.  In order to exploit the full potential of $\ttbar + \etmiss$ searches it is thus crucial to refine the modelling of the $\etmiss$ and $m_\mathrm{T2} $ distributions  in $\ttbar Z$ production. We have furthermore found that with of $300 \, {\rm fb}^{-1}$ data and under the assumption that the SM backgrounds are known to 20\%, it should be possible to distinguish between the two CP hypotheses for mediator masses up to around $200 \, {\rm GeV}$ in all scenarios that lead to $\mu \lesssim 1$. Finally, we have presented 95\% CL exclusion bounds in the $M$--$\hspace{0.25mm} g_t$  and $M$--$\hspace{0.5mm} \tan \beta$ planes which allow to illustrate the LHC reach in the context of the DMF spin-0 simplified models  and 2HDMII plus singlet extensions, respectively. The upshot of our recast is that spin-0 simplified models that lead to an effective coupling strength of order 1 of the mediators to top-quark pairs can typically be tested  for masses up to (or even above) the $\ttbar$ threshold at the HL-LHC. 

The DM top-quark couplings of spin-0 $s$-channel simplified models can be probed at the LHC also in other channels such as $pp \to {\rm jets} + \etmiss$ and $pp \to \ttbar$.   In both cases it should be possible to use angular correlations to either enhance the sensitivity of these searches or to determine the CP properties of the involved mediators. Since the $pp \to \ttbar + \etmiss$, $pp \to  {\rm jets} + \etmiss$ and $pp \to \ttbar$ probe different aspects of the underlying theory, we believe that it is crucial to combine all available information on the DM top-quark couplings  to fully exploit the LHC potential. We look forward to further investigations in this direction.

\acknowledgments 
We thank Giorgio Busoni for spotting a sign mistake in (\ref{eq:2HDM}) of the first version of this article. UH thanks the CERN Theoretical Physics Department for continued hospitality and support. PP acknowledges access to the LHC Computing Grid at Nikhef.

\providecommand{\href}[2]{#2}\begingroup\raggedright\endgroup

\end{document}